# Identification of Solid-Electrolyte Interphase Species by Joint Characterization of Li-ion Battery Chemistry by Mass Spectrometry and Electro-Chemical Reaction Networks

Mona Abdelgaid,[1,2] Oliver Hvidsten,[2,3] Théo Sombret,[4-7] Egon Kherchiche,[4-7] Julien Maillard,[5,7] Antonin Gajan,[6] Patrick Bernard,[6] Kamila Kazmierczak,[8] Mauricio Araya-Polo,[9] Germain Salvato Vallverdu,[7,10] Carlos Afonso,[4,7] Pierre Giusti,[4,5,7] Kristin A. Persson*[1-3]

1. Bakar Institute of Digital Materials for the Planet, University of California at Berkeley, Berkeley, CA, 94720, United States.
2. Materials Science Division, Lawrence Berkeley National Laboratory, Berkeley, CA, 94720, United States.
3. Department of Materials Science and Engineering, University of California at Berkeley, Berkeley, CA, 94720, United States.
4. Université de Rouen Normandie, INSA Rouen Normandie, Université de Caen Normandie, ENSICAEN, CNRS, Institut CARMeN UMR 6064, F-76821 Mont-Saint-Aignan Cedex, France.
5. TotalEnergies OneTech, TotalEnergies Research & Technology Gonfreville, BP 27, 76700 Harfleur, France.
6. Saft, Corporate Research, 33074 Bordeaux, France.
7. International Joint Laboratory-Complex Matrices Molecular Characterization (iC2MC), TRTG, BP 27, 76700 Harfleur, France.
8. TotalEnergies OneTech Belgium, 7181 Seneffe, Belgium.
9. TotalEnergies EP Research & Technology, Houston, TX, 77002, United States.
10. Universite de Pau et des Pays de l'Adour, CNRS, IPREM, 64000 Pau, France.

*Corresponding author

E-mail: kristinpersson@berkeley.edu

**Abstract:**

The formation and stability of the solid electrolyte interphase (SEI) play a central role in determining the long-term performance and safety of modern electrochemical energy storage systems. Despite decades of research, the SEI's heterogeneous, dynamic, and multi-phase nature has defied comprehensive molecular-level characterization, creating a critical knowledge gap that limits rational battery design. In this work, we introduce a computational-experimental framework that integrates high-throughput quantum chemistry calculations, data-driven electro-chemical reaction networks (eCRNs), stochastic algorithms, and Laser Desorption/Ionization Fourier Transform Ion Cyclotron Resonance Mass Spectrometry (LDI-FTICR-MS) to unravel SEI formation in carbonate-based electrolytes without imposing predefined mechanisms. We constructed the most comprehensive eCRN to date, spanning over 10,000 species and 209 million reactions. Through stochastic network analysis, we successfully recovered 27 species that were previously reported in literature and predicted 28 novel SEI species—nearly doubling our scientific knowledge in this area. Each new species was rigorously confirmed through advanced mass spectra analysis of its distinct molecular and isotopic signatures. We kinetically refined the formation pathways for a select set of both previously reported and novel SEI products, revealing kinetically feasible elementary reaction mechanisms with activation barriers below 1 eV. This computational-experimental approach deepens our molecular-level understanding of SEI chemistry and supports the rational design of advanced electrolytes and engineered interphases for next-generation lithium-based batteries.

## Introduction:

Rechargeable batteries form the backbone of contemporary energy storage technologies, and lithium-ion batteries in particular power devices ranging from portable electronics to electric vehicles and large-scale energy systems.[1,2] However, pushing the boundaries of electrochemical energy storage performance, safety, and longevity demands a deeper understanding of the diverse set of reactions that occur at the electrode-electrolyte interface, many of which contribute to the formation of the solid electrolyte interphase (SEI). The SEI was introduced by Peled[5] in 1979 as an electronically insulating and ionically conducting passivation layer, formed as a result of electrolyte reduction under the highly reducing conditions near the anode during the first cycles.[6,7] Without the SEI, uncontrolled electrolyte breakdown reactions lead to rapid capacity loss, poor Coulombic efficiency, and eventual battery failure.[7] Early models offered a biphasic view of the SEI, comprising of a stable inorganic inner layer in contact with the anode, and a more organic, solvent-rich outer layer facing the electrolyte.[8] Until today, the SEI continues to be considered "the most important but least understood (component) in rechargeable Li-ion batteries"[9] due to the complex nature of the spontaneous chemical and electrochemical processes involved in its formation and the lack of adequate direct characterization of its physical properties.[6,10] Its structure, morphology, composition, and properties are highly sensitive to many factors, including electrolyte components, electrode material, solvents, additives, temperature, and areal current.[11]

Historically, much of what is known about the SEI has come from bulk or ensemble-averaged analytical methods such as X-ray Photoelectron Spectroscopy (XPS),[12] Scanning Electron Microscopy (SEM),[13] X-ray Diffraction (XRD),[13] Fourier Transform Infrared Spectroscopy (FTIR),[14] Time-of-Flight Secondary Ion Mass Spectrometry (ToF-SIMS),[15] and others.[16,17] While these analytical techniques have been essential in shaping our current understanding, they each provide partial insight into the SEI's complex structure and chemistry. Electron microscopy and diffraction-based approaches like SEM and XRD excel at revealing surface morphology, cross-sectional structure, and crystalline phase information, yet they face significant challenges in directly probing the chemical bonding environment and molecular-level interactions within the SEI.[18,19,13] XPS, on the other hand, has been widely used to analyze surface composition and identify key SEI species (e.g. lithium fluoride (LiF), lithium carbonate ($Li_2CO_3$), and organic compounds) with nanoscale spatial resolution.[17,20] Despite its sensitivity, the poor electronic conductivity inherent to the SEI creates substantial difficulties during measurement, as charging effects during standard XPS analysis result in binding energy shifts that are notoriously difficult to calibrate and correct.[17,18,20] Moreover, the coexistence of multiple lithium-containing compounds and multilayer SEI architecture that vary spatially across the electrode surface further limit the accuracy and interpretability of these traditional approaches. Importantly, distinguishing organic from inorganic layers presents acute analytical difficulties. FTIR spectroscopy has become the primary tool for evaluating the organic SEI and electrolyte chemistry due to the many IR-active vibrational modes of both salts and carbonate solvent comprising the electrolyte and its decomposition products.[14] However, FTIR analysis faces significant practical limitations including inherent sensitivity constraints, complex sample handling requirements, and susceptibility to environmental interference.[16,17]

To precisely determine the molecular composition of the SEI on the anode surface, advanced high-resolution mass spectrometry, particularly Fourier Transform Ion Cyclotron Resonance Mass Spectrometry (FTICR-MS), offers a powerful approach.[21–24] This technique provides exceptional mass resolution, capable of distinguishing between species that differ by less than the mass of an electron,[24] and delivers mass accuracy sufficient to assign distinct molecular formulas to each detected signal.[23] Such precision makes it particularly effective for identifying unknown compounds in complex chemical systems.[23] FTICR-MS also features a broad dynamic range, allowing the detection of trace-level species and isotopic variants critical for confident molecular identification.[25] When combined with Laser Desorption/Ionization (LDI), it further enables spatially resolved surface analysis, facilitating detailed mapping of the SEI's molecular heterogeneity.[26,27]

Recent studies have demonstrated the remarkable capabilities of FTICR-MS in advancing molecular-level insight into SEI and electrolyte chemistry. Using matrix-assisted FTICR-MS and TOF-MS, Sombret *et al.* monitored the degradation behavior of various lithium salts under controlled and air-exposed conditions, showing that electrolytes in the solid state remained chemically stable for over 24 hours, in contrast to rapid degradation observed in solution.[28] Maillard *et al.* utilized LDI-FTICR-MS method to analyze SEI composition and identified degradation products from both salt and solvent decomposition.[27] The high elemental diversity (C, H, O, Li, P, F) led to over 4,000 signals per electrode in each ionization mode, making manual assignment extremely challenging.[27] This highlights the need for combining high-quality characterization with data-driven approaches that leverage first-principles calculations and computational electrochemical reaction networks (eCRNs) to aid in the identification of SEI components.[29–31]

To bridge this gap, Persson and co-workers developed the first scalable computational frameworks to explore complex electrochemical reaction networks.[29–32] Xie *et al.* applied graph-based models and shortest-path algorithms to identify the formation pathways for key SEI products.[29] Barter *et al.* introduced the High-Performance Reaction Generation method, which leverages filter-based reaction enumeration and stochastic analysis, to recover known SEI products and identify previously unrecognized SEI species.[30] Using kinetic Monte Carlo (kMC) simulations informed by automated eCRN analysis and ab initio calculations, Spotte-Smith *et al.* explored SEI formation and evolution of the solvent decomposition, revealing the mechanistic origins of the Peled-like bilayer interphase structure on the negative electrode of Li-ion batteries.[31] Such kMC approaches are increasingly recognized as essential tools for bridging molecular-scale phenomena with macroscopic battery models, offering the computational accuracy needed to understand complex electrochemical reactions and evolving solid interfaces.[33] By combining first-principles and data-driven simulations with microkinetic models, it is now possible to rapidly generate hypotheses for experimental characterization and in-depth mechanistic studies of complex reactive processes. However, notably, no eCRN has yet been constructed that simultaneously captures both salt and solvent decomposition, along with their joint products. This limitation has prevented the data-driven identification of hybrid species such as fluorinated organophosphate-carbonates that contain both phosphorus/fluorine atoms from salt fragments and carbonate functionalities from solvent fragments.

In this work, we combine LDI-FTICR-MS with quantum chemistry calculations, data-driven electrochemical reaction networks, and stochastic simulations to characterize an SEI formed on a graphite anode in a typical carbonate-based LIB electrolyte, as shown in **Figure 1**. The ultra-high resolution LDI-FTICR-MS technique identified over 30,000 signals for the formation cycle for $S/N > 3$, which include multiple isotopic and ionic variants of key SEI species. Bridging this experimental richness with theoretical prediction, we constructed and analyzed a massive-scale electrochemical reaction network, including both salt and solvent reactivity, to systematically generate candidate SEI species for direct comparison with LDI-FTICR-MS. We recovered 27 known SEI components and identified 28 novel species previously unreported in the literature. Additionally, we ide initial state, thereby ensuring representation ntified kinetically feasible pathways for the formation of a representative set of these novel SEI species. Together, this study reinforces the critical advantage of coupling theoretical and experimental approaches to characterize the SEI. By doing so, it paves the way toward the rational design of electrolytes and engineered interphases for next-generation battery technologies.

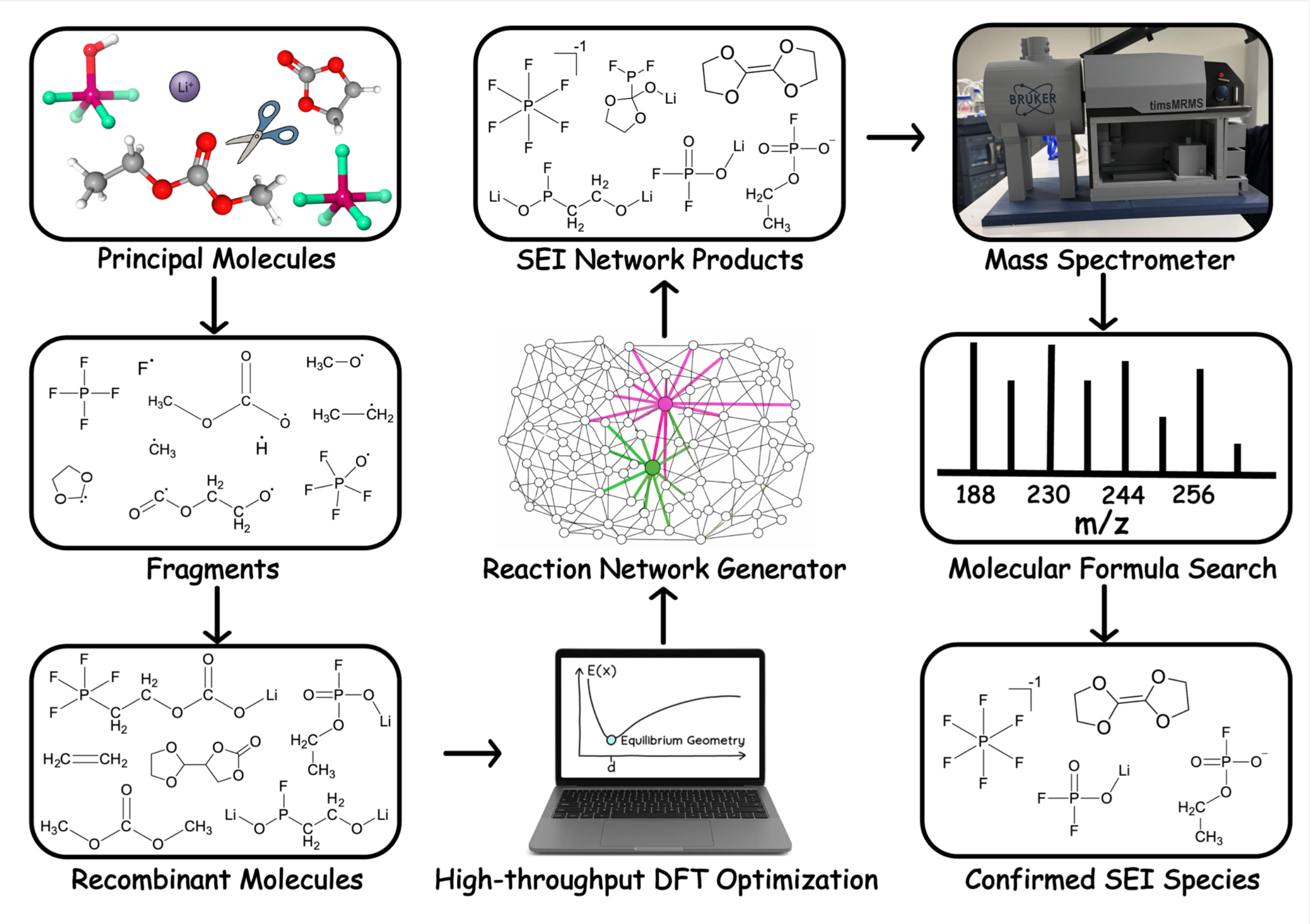

**Figure 1**. Workflow illustrating the fragmentation of the principal molecules and further recombination of these fragments to form new, larger molecules. All molecules are then analyzed using high-throughput DFT calculations, which provides an understanding of their 3D geometry, thermodynamic quantities (e.g. energy, enthalpy, and entropy), and vibrational frequencies. These properties are then filtered and enumerated to generate electrochemical reaction networks and stochastically identify key network species using a set of heuristics. Then, the computationally

identified species are experimentally confirmed through mass spectra analysis of molecular and isotopic signatures.

**Experimental methods:**

**Electrolyte preparation:**

The study was carried out on negative electrode coming from batteries filed with an electrolyte composed of highly pure $LiPF_6$ (purity > 99.5 %, supplied by Soulbrain), and battery grade solvents ethyl carbonate (EC) and ethyl methyl carbonate (EMC) (both provided by Capchem, purity > 99.5 %): 1 M $LiPF_6$ in EC/EMC (30/70) vol. The electrolyte was prepared under an argon atmosphere in a glove box and stored in airtight aluminum containers. All compounds used for the electrolyte formulation were placed in a drying oven overnight. Before use, the spatulas were cleaned with dimethyl carbonate (DMC, > 99.5% pure, supplied by Capchem). After formulation, the electrolyte was left in the glove box overnight before using to ensure the complete dissolution of the lithium salts.

**Assembling, electrochemical tests, and dismantling of the cells:**

Single layer pouch cells were assembled in a dry room (dew point below −45 °C), with a negative electrode of 2.74x2.74 $cm^2$ and a positive electrode of 2.54x2.54 $cm^2$. The active materials of the negative and positive electrodes were graphite and a lithiated lamellar oxide $LiNi_{0.6}Mn_{0.2}Co_{0.2}O_2$ (NMC 622), respectively. The electrodes were all provided by Saft company. A Celgard 2325 type separator was placed between the two electrodes. After assembly, the cells were left to dry at 80 °C in a vacuum chamber for at least one night. Once dried, the pouch cells were filled with electrolyte and sealed in a glove box before cycling. Electrochemical analyses were performed using a BioLogic BCS-805 battery cycler. All the cycles were performed using a constant current mode between 2.7 V and 4.2 V. All the cells only underwent a cycle at a C-rate of C/20 - D/20 at 60 °C. The objective of this cycle is to study the initial formation of SEI. The cells were disconnected immediately after this first cycle. Pouch cells were then recovered and dismantled in an Ar-filled glove box. The electrodes were washed in EMC two times in cups of aluminum for 30 seconds. The objective was to remove, before analysis, the Li-salt possibly trapped in the electrode pores. Once washed, the electrodes were dried at room temperature in the glove box before further handling. The porous separator containing the electrolyte was also placed in a few milliliters of EMC solvent to recover the degraded electrolyte trapped in the pores of the separator. All the electrochemical data are presented in **Figure S1** and **Table S1**.

**Mass spectrometry procedure**

Mass spectrometry analyses were conducted using a FTICR-MS equipped with an 18 T superconducting magnet (MRMS, Bruker Daltonics, Bremen, Germany) equipped with a laser desorption ionization (LDI) source with a Nd:YAG × 3 laser at 355 nm. Prior to analysis, the electrodes were attached to a stainless steel MALDI plate with conductive aluminum adhesive tape. All sample handling, including electrode preparation and deposition, was performed under argon atmosphere in a glove box ($O_2$, $H_2O$ < 0.8 ppm) to avoid oxidation or moisture contamination. The samples were transferred from the glovebox to the mass spectrometer source in a plastic zip bag filled with $N_2$ gas to avoid contact with air.[27] The mass spectra were acquired in positive mode in the *m/z* range 90-1000. Conditions were optimized to afford the transmission of low *m/z* ions: plate offset 100 V; deflector plate 220 V; Funnel 1 RF 150 $V_{pp}$, Funnel 2 RF 150 $V_{pp}$; multipole RF 200 $V_{pp}$; TOF 0.500 ms; collision cell RF 650.0 $V_{pp}$, RF amplitude 350 $V_{pp}$, RF frequency 6 MHz, chirp excitation power 25 % and number of scans 50. The source parameters were set up as followed: laser frequency 10,000 Hz, laser focus small, laser power 35%, number of shots by burst 680. The analog time domain signal was digitalized using 8 million data points, apodised and zero filled once before Fourier transform. With a low acquisition mass of *m/z* 90, this resulted in a time domain signal of 1.4 s yielding a mass spectrum with a resolving power of 700,000 at *m/z* 200 (magnitude mode and 1ω detection). The analyses were performed on four graphite negative electrodes after the formation cycle. The mass spectra are presented in **Figure S2**.

Molecular formula and data processing were performed using DataAnalysis v 6.1 (Bruker). Mass spectra were internally calibrated with carbon cluster cations coming from the graphite electrode surface. The molecular formula assignment of the species was realized with the SmartFomula tool. A tolerance of 0.5 ppm was allowed to the identification with the following elemental constraints $C_{0\text{-}50}$, $H_{0\text{-}100}$, $O_{0\text{-}20}$, $N_0$, $F_{1\text{-}5}$, $Li_{1\text{-}5}$, $P_{1\text{-}5}$. In many cases, owing to the high number of elements to consider, several molecular formulas were obtained with the 0.5 ppm mass tolerance.

Discrimination between them was carried out by considering the isotopic fine structure with the mSigma score given by the DataAnalysis SmartFomula tool and manual control.

**Computational methods:**

**Data generation through fragmentation and recombination**

The set of possible species that could contribute to the SEI formation in LIBs was generated by fragmenting the principal molecules, including the solvent molecules (e.g. EC and EMC), salts (e.g. $LiPF_6$), and one of the early salt decomposition products (e.g. $PF_4OH$). Importantly, $PF_4OH$ was included in the initial state, thereby ensuring representation of concerted salt decomposition mechanisms such as the $PF_5$-$Li_2CO_3$ adduct dissociation to LiF, $CO_2$, and $LiPOF_4$.[34] This approach ensures that the combined chemical space spanning both salt-derived (P, F) and solvent-derived (C, H, O) fragments is accessible during recombination, enabling the formation of hybrid species that would be impossible through separate decomposition pathways. We implemented a systematic two-step fragmentation approach. The first step involved systematically breaking bonds within each principal molecule to generate an initial set of molecular fragments. In the second step, these fragments were further fragmented, producing a broader and more diverse pool of smaller molecular units. Following this hierarchical fragmentation, these fragments were allowed to selectively recombine to form larger molecules. This combination of sequential fragmentation and strategic recombination yielded a chemically diverse set of species that could serve as intermediates connecting original electrolyte components to final SEI products. All principal molecules, fragments, and recombinant molecules were optimized using high-throughput Density Functional Theory (DFT) calculations, which provides an understanding of their 3D geometry, thermodynamic quantities (including total electronic energy, enthalpy, entropy, and Gibbs free energy), and vibrational frequencies. Following molecule generation and analysis, the resulting dataset was then augmented with molecules from the lithium-ion battery electrolyte dataset[35] that fell into three categories: those containing C, H, O, and/or Li; those containing P, F, and/or Li or H; and those containing P, F, C, H, O, and/or Li. This workflow enabled a comprehensive exploration of the complex reactive landscape governing LIB electrolyte chemistry.

**Density functional theory**

DFT calculations were performed using the Q-Chem electronic structure package.[36] Geometry optimizations for all species within the reaction network utilized the range-separated hybrid functional ωB97X-V[37] in conjunction with the triple-zeta def2-TZVPPD[38] basis set, which incorporates diffuse and polarization functions for improved accuracy in non-covalent interactions. Solvation effects were incorporated implicitly via the SMD continuum solvation model,[39] parameterized to emulate a lithium-ion battery electrolyte environment. A dielectric constant of $\varepsilon = 18.5$ was selected to approximate the permittivity of a 3:7 (v/v) EC: EMC solvent mixture, while remaining solvent descriptors (e.g. surface tension, atomic radii) were retained at values corresponding to pure ethylene carbonate to maintain consistency with prior studies.[35]

Transition state (TS) structures were initially located using the automated AutoTS[40] protocol within the Jaguar[41] software package, employing the ωB97X-D functional[42] paired with the split-valence def2-SVPD basis set[38,43] and the C-PCM[44] implicit solvation model to balance computational efficiency and reliability for saddle-point searches. TSs that were unable to be identified via this procedure were manually estimated and optimized by Jaguar's built-in TS optimization methods. To ensure robust characterization, all TS candidates were rigorously validated by (1) confirming the presence of a single imaginary vibrational frequency corresponding to the reaction coordinate and (2) performing intrinsic reaction coordinate (IRC)[45] analyses to verify connectivity between reactant and product minima. Final TS geometries and energies were recalculated at the higher-level ωB97X-V/def2-TZVPD/COSMO[46] theory to provide more accurate reaction energy profiles.

**Electrochemical reaction network generation**

To construct the electrochemical reaction network, we first applied a filtering methodology.[30] Starting with a dataset of 23,625 candidate species, we applied a series of systematic filters to eliminate species that are chemically unreasonable or undesirable under the conditions studied. The framework then enumerates all stoichiometrically valid reactions and employs user-defined criteria to eliminate reactions based on physical or practical criteria while preserving a chemically diverse and meaningful reaction network. A detailed description of species and reaction filters employed in this work can be found in Spotte-Smith *et al.*[30]

**Stochastic simulations**

We employed the Reaction Network Monte Carlo (RNMC) framework developed by Zichi *et al.*,[47] which leverages a scalable, parallel implementation of Gillespie's direct method[48] with significant optimizations for handling massive reaction networks. Specifically, we employ kMC simulations as a stochastic sampling framework to identify which reactions occur and which species form via multiple reaction pathways, emerge most readily, and persist without being consumed. We do not attempt to resolve reaction timescales or quantitative product distributions. Building on that foundation, we performed 50,000 kMC simulations in parallel under 0.0 V vs. $Li/Li^+$ and 298.15 K, designed to probe the stochastic evolution of complex reactive systems. In order to efficiently estimate the system's species dynamics, each simulation was initialized with a representative electrolyte composition of 30 $Li^+$, 30 $PF_5$, 30 $PF_4OH$, and a solvent mixture of 30 molecules split between EC and EMC in a 30:70 volumetric ratio (12 EC and 18 EMC). In addition, the system is treated within the grand-canonical ensemble, with the anode acting as an electron reservoir, enabling control of the voltage via the electron chemical potential. $PF_4OH$ was included in the initial state to circumvent reaction filters that excludes reactions involving more than two simultaneous bond breakages/formations, thereby ensuring representation of concerted salt decomposition mechanisms such as the $PF_5$-$Li_2CO_3$ adduct dissociation to LiF, $CO_2$, and $LiPOF_4$.[34] In addition, more computationally intensive simulations were initialized with a "true-to-experiment" composition of 10,908 molecules of EC, 16,395 molecules of EMC, and 2,697 molecules of the salt. This represents a volume of 4500 $nm^3$ with an approximate 1 M of $LiPF_6$ concentration in 30:70 EC:EMC. For the smaller electrolyte system, 30 molecules of each major component were sufficient to adequately sample the system's reactive behavior while maintaining computational tractability, providing statistically meaningful results through the ensemble of thousands of parallel simulations.

In these kMC simulations, the system evolved stochastically from the defined initial state, with reaction events selected probabilistically based on the supplied rate coefficients. Given the scale of the reaction network which entails millions of reactions, it is computationally infeasible to assign accurate, individually calculated rate constants. To address this, we employed uniform rate coefficients: all unimolecular reactions were assigned a constant rate of $k_0$, while bimolecular reactions were assigned a rate of $k_0/V$, where V is the system volume introduced to maintain correct dimensionality. Specifically, we calculated rate constants using transition state theory:

$$k = \frac{k_B T}{h} \cdot e^{\left(\frac{\Delta G^{\ddagger}}{k_B T}\right)}$$

where $k_B$ is Boltzmann constant, $h$ is Planck constant, $T$ is temperature, and $G^{\ddagger}$ is the activation energy barrier. We used a systematic barrier construction where $\Delta G^{\ddagger}$ = constant barrier for exergonic reactions and $\Delta G^{\ddagger} = \Delta G_{reaction}$ + constant barrier for endergonic reactions. This approach enables consistent and tractable sampling of the reaction network while preserving the relative dynamics between uni- and bimolecular processes. With sufficient sampling, all reactions represented in the network that would occur under accurate kinetics will be captured in kMC simulations with fixed rate coefficients. Previous computational studies of chemical reaction networks relied on simplified approaches, such as Stocker *et al.* combustion network analysis, which utilized reaction thermodynamics coupled with arbitrary energy barriers to explore network dynamics.[49] Additionally, Spotte-Smith *et al.* successfully recovered experimentally observed SEI products without knowledge of reaction kinetics, showing that reaction thermodynamics alone can be sufficient to predict reasonable reaction pathways as well as SEI network products.[30] Furthermore, Blau *et al.* demonstrated on a 6,000-species eCRN that thermodynamic-driven analysis can autonomously identify viable reaction pathways, recovering all known formation mechanisms for lithium ethylene dicarbonate while discovering three additional novel routes.[32]

## Results & Discussion

We generated a comprehensive eCRN aimed at capturing the formation of the SEI in LIBs. The initial dataset comprises 23,625 species with varying charges (−1, 0, +1) and spin multiplicities (1, 2, 3), as shown in **Figure 2**. The dataset was biased towards the C-H-O-Li chemical system because most principal molecules are organic in nature, with 4,053 phosphate-containing molecules present as well. The bonding motifs observed for phosphorus were limited (only F-P, O-P, while a small number of C-P, H-P, and Li-P bonds were also present) because these molecules are all derived from salt and solvent joint decomposition products. Furthermore, we examined the molecular size distribution based on electron count and observed that while the majority of species are relatively small molecules, the dataset includes a subset of considerably relatively larger molecules compared to the original electrolyte solvent molecules and salt anions, containing up to 120 electrons. While the dataset is intentionally biased toward smaller molecular species (e.g. only one recombination step of the molecular fragments) due to the high computational cost of larger molecules, it effectively addresses the core chemical processes governing SEI formation and provides a robust foundation for future extensions to larger molecular systems.

To refine the dataset, we implemented a number of user-defined filters (summarized in the supporting information) to remove undesirable species, yielding a set of 11,867 filtered species. These filtered species were then organized into composition-based groups, where each group contains entries consisting of either a single molecule or a pair of molecules whose combined composition matches that of the group. Across all groups, we generated every stoichiometrically valid reaction and filtered them using physical and practical criteria, yielding 208,929,658 reactions—the largest eCRN for LIB SEI formation to date (**Figure S3**) and the first to include coupled salt–solvent decomposition pathways.

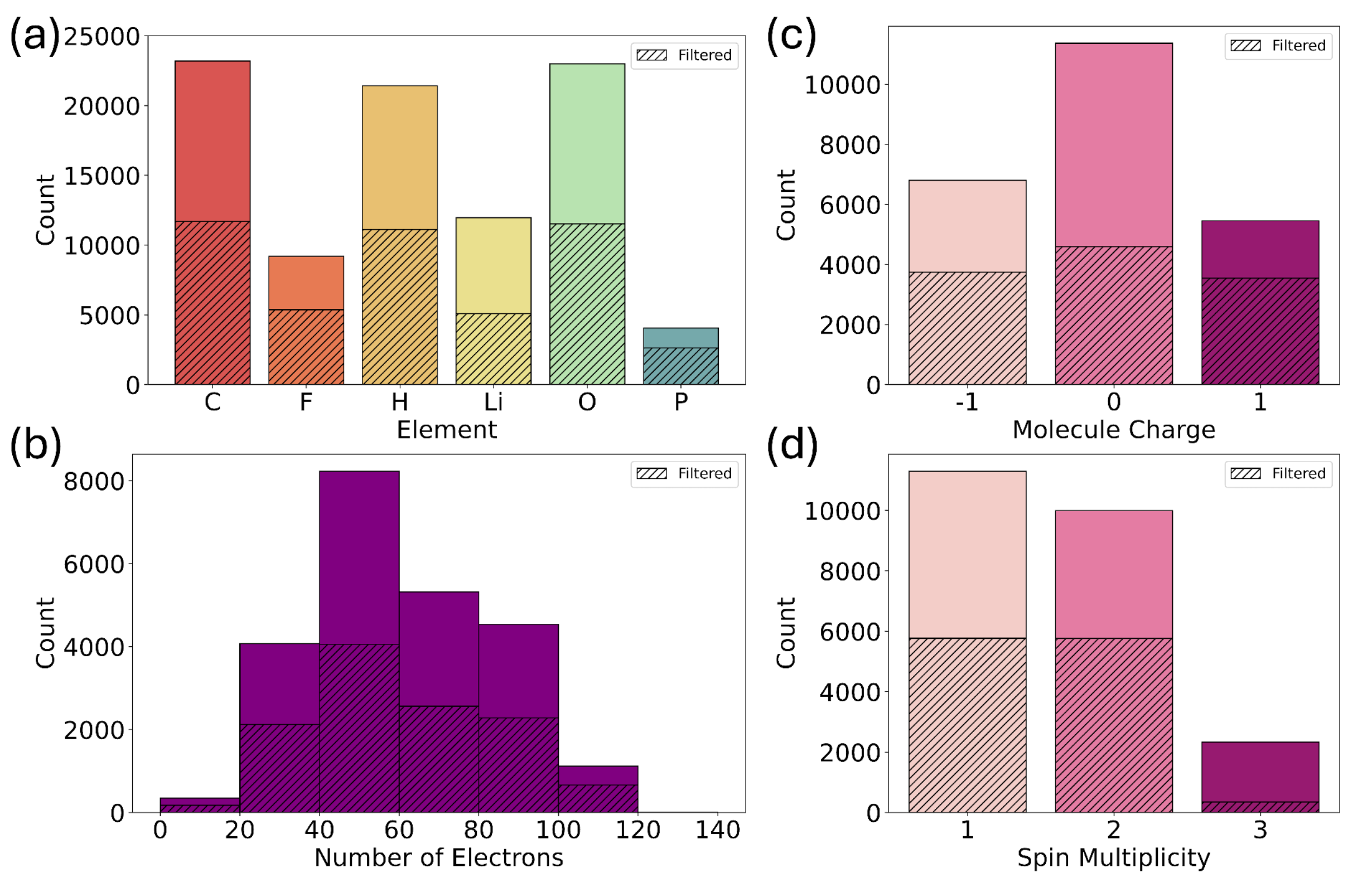

**Figure 2**. An analysis of the composition of the dataset used to construct the eCRN. Each plot includes a solid-colored bar representing counts in the original dataset fed to the filtering framework, and a hatched overlay bar for the filtered subset. (a) Number of unique molecules containing each elemental species; (b) distribution of molecules by number of electrons; (c) number of molecules with formal charges of −1, 0, and +1; and (d) number of molecules with spin multiplicity values of 1, 2, and 3.

To analyze the eCRN, we performed thousands of parallel kMC simulations, with each simulation producing distinct reaction trajectories that collectively map the complex reactive landscape of the SEI formation. These trajectories enable two complementary analysis approaches, including targeted pathway discovery and autonomous product identification. For known species of interest, we trace trajectories to identify the shortest reaction pathways leading to their formation and rank these pathways using cost functions. For unbiased product discovery, we apply heuristic criteria across all trajectories to identify likely network products. Species qualify as network products if they: (i) are formed substantially more than they are consumed (ratio of producing reactions to consuming reactions > 1), (ii) cumulate to a significant degree on average (average final concentration must be > 0.1), and (iii) are accessible through low-cost pathways (number of elementary steps from initial species to final product < 15). It is important to note that the identified network products do not necessarily represent the metastable or stable species that would be observed experimentally in the SEI, nor are they necessarily exhaustive, as they depend on which species are included in the eCRN construction, the chosen threshold parameters, and initial conditions of the kMC simulations. Rather, the network products serve as valuable hypotheses for species that might form in actual reactive systems, which can subsequently be validated experimentally using techniques such as ultra-high resolution LDI-FTICR-MS.

The utility of our approach was evaluated through the analysis of a diverse set of network products (predominantly neutral molecules with a smaller subset of anions) generated under the various initial kMC conditions, as described in the methodology section. To search for these molecules in LDI-FTICR-MS spectra in the positive and negative ionization modes, we developed a Python code that systematically generates all relevant ionic forms for each molecule. For the positive ionization mode, network products were cationized by adding $Li^+$ or $H^+$, while in the negative mode, we searched for their deprotonated and $Li^+$-deficient counterparts. The algorithm incorporates common adduct formations by combining the base molecular formula of the monomer species with an adduct (LiF or $Li_2CO_3$) followed by ionization with either $Li^+$ or $H^+$. Additionally, we generated corresponding dimeric and trimeric species to account for potential aggregation effects. For every candidate ion, we calculated the exact mass of both the parent ion and its isotopic variants. Species confirmation requires the detection of signal corresponding to both the parent ion and its expected isotopic peaks in LDI-FTICR-MS. This systematic ion generation and matching pipeline enabled robust cross-checking between the reaction network predictions and LDI-FTICR-MS, revealing several previously unreported species (**Table S2**).

Our computational approach recovered 27 SEI products that are previously reported in literature (**Figure 3**). These include (i) gases such as $H_2$, $C_2H_4$, and CO;[50] (ii) inorganic species including inorganic carbonates (e.g. $Li_2CO_3$), lithium oxalate ($Li_2C_2O_4$), and LiF;[51–53] (iii) organic species like lithium methyl carbonate (LMC),[54,55] lithium ethyl carbonate (LEC),[55] lithium butylene dicarbonate (LBDC),[52] dimethyl 2,5- dioxahexane carboxylate (DMDOHC),[56] to name a few; and (iv) phosphate compounds[27,34] such as hexafluorophosphate ($PF_6^-$), lithium phosphorodifluoridate ($LiPO_2F_2$), lithium phosphorofluoridate ($Li_2PO_3F$), phosphoric acid ($H_3PO_4$), and others. It should be noted that LDI-FTICR-MS has inherent limitations in detecting very low molecular weight and volatile species (e.g. CO, $C_2H_2$, $Li_2CO_3$, etc) due to their poor retention during the desorption process and the low mass detection limits of the technique. Overall, the successful recovery of these well-known SEI species demonstrates the robustness of our computational approach, particularly considering that reaction kinetics are entirely ignored in network exploration.

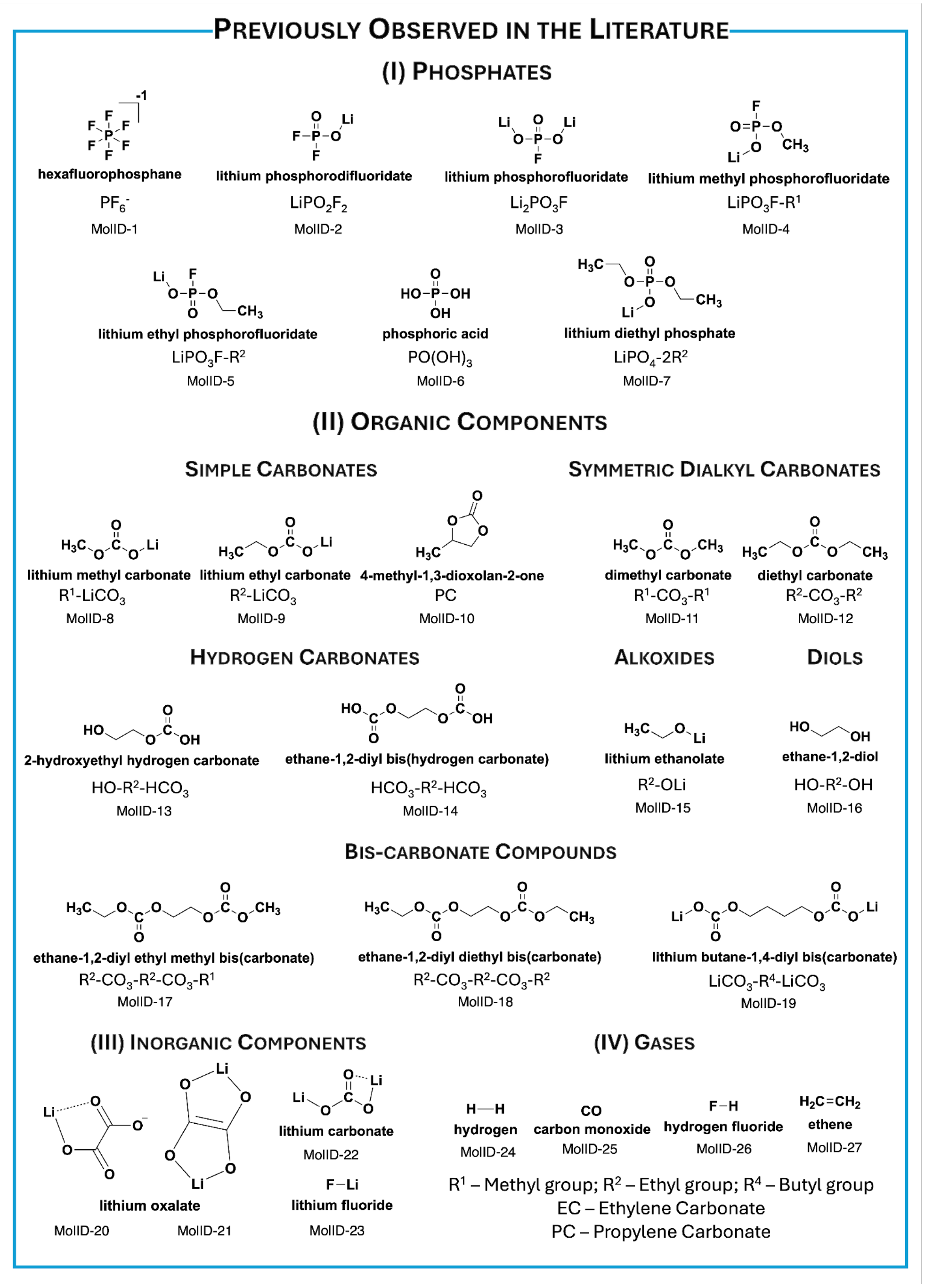

**Figure 3**. Stochastic network analysis recovered 27 species which have been reported in the SEI literature. These species include major gases, inorganic molecules, linear alkyl carbonates, and phosphate derivatives.

Most remarkably, our computational framework uncovered several novel products that have not previously been proposed to participate in SEI formation (**Figure 4**). These molecules were further experimentally confirmed through mass spectral analysis of molecular and isotopic signatures, as depicted in **Figure 5**. These 28 unique species include fluorinated organophosphate-carbonates, cyclic and bicyclic alkyl carbonates, oxalates, formates, diols, ethers, aldehydes, carboxylates, and dioxolanes. These newly identified species have known properties that could support the overall mechanical, chemical, and transport properties of the SEI. Mechanical robustness is critical for an SEI as cracking reactivates passive surfaces, significantly damaging the overall lifetime and performance of the battery.[57] Oxalates (e.g. MolID-43) can decompose preferentially during the initial battery cycles to form $Li_2C_2O_4$ which contributes to a compact, ionically conductive SEI layer with enhanced mechanical stability and reduced electrolyte decomposition.[58–60] Vinyl-bearing formates (e.g. MolID-47 and MolID-48) can undergo reductive polymerization during the initial charge cycle, forming a flexible, cross-linked polymer network within the SEI.[61–65] These vinyl-initiated radical polymerization pathways are known to generate cohesive, polymer-rich layers that enhance SEI elasticity, relieve mechanical stress, suppress dendrite growth, improve ionic conductivity, and effectively block electron tunneling. [61–65] Cyclic carbonates, particularly 1,3-dioxolane derivatives (e.g. MolID-35 to MolID-39), also undergo polymerization and decomposition to form hybrid organic-inorganic SEI layers with enhanced mechanical stability, allowing the interphase to accommodate electrode volume changes and mitigate crack propagation during lithiation/delithiation cycles.[66] The longer-chain alkyl carbonates (e.g. MolID-28 to MolID-34) and flexible ether molecules (e.g. MolID-46) could provide enhanced mechanical accommodation in SEI layers through their increased conformational flexibility. Moreover, diols (e.g. MolID-50) can act as crosslinking agents, enhancing the mechanical strength of the SEI layers in LIBs.[67,68] Chemical stability is another essential performance parameter, especially given the SEI's primary role is surface passivation of the anode.[69] The fluorinated phosphate-carbonate hybrids[66,70,71] (e.g. MolID-54) may assist in the stabilization between the organic and inorganic components of the SEI due to their intermedial character while the aldehydes (e.g. MolID-44 and MolID-45) and carboxylates[72,73] (e.g. MolID-41 and MolID-42) have an overall chemically stabilizing affect that assists the passivation of the interphase with its liquid environment. The final performance parameter is the overall transport of $Li^+$ ions through the SEI during lithiation/delithiation cycles. The identity of the paramount Li conduction component of the SEI is still under debate,[74] however, recent work point to the importance of mixed anion amorphous phases within the Li-O-P-F chemical space.[74,75] In particular, our recent computational study demonstrated that bulk amorphous $LiPO_2F_2$ (MolID-2) exhibits fast ionic conduction with exceptionally low Li-interstitial defect formation energies and rapid $Li^+$ diffusion, suggesting that mixed-anion Li-O-P-F phases can serve as primary Li-conducting channels within the SEI matrix.[74] Such regions within the SEI may be promoted by fluorinated phosphate-carbonate hybrids,[66,70,71] hydrogen bonding networks created by diols, and the polar groups found in ethers and linear carbonates. While future studies are necessary to confirm the role of each of these species in the SEI, this diverse identification provides data on the chemical make-up by which the overall interphase function is orchestrated. Additionally, the information broadly offers new molecular insights that could inform future design principles for tailoring electrolyte formulations that yield more stable, ionically conductive, and self-healing interphases. For instance, electrolyte additives containing vinyl functionalities can be designed to decompose into the identified vinyl-bearing formates that enhance mechanical flexibility, while formulations promoting coupled salt-solvent pathways can target the formation of fluorinated phosphate-carbonate hybrids that optimize ionic transport. Rather than empirical optimization, electrolyte chemists can now work backwards from desired SEI properties to identify the molecular precursors and reaction conditions needed to generate specific beneficial components, transforming electrolyte development from trial-and-error to knowledge-driven molecular engineering.

As stated earlier, these molecules were validated through exact mass matching with LDI-FTICR-MS, achieving mass accuracy to the 4$^{th}$ and 5$^{th}$ decimal places for species with molecular weights below and above *m/z* 175, respectively. While this high-precision mass matching provides strong evidence for molecular presence, we acknowledge that mass spectrometry alone cannot definitively distinguish between structural isomers or confirm molecular connectivity (**Figure S4**). This study successfully demonstrates a joint computational-experimental methodology, with detailed structural validation representing an important avenue for future research using complementary analytical techniques. Furthermore, our eCRN models decomposition products to a defined computational depth, recognizing that the observed species (e.g. cyclic/bicyclic carbonates, and other organic fragments) may undergo further reactions, including bond breaking, ring-opening, polymerization, crosslinking, etc, that extend beyond the current network boundaries. While our computational framework is limited to smaller molecules, experimental 2D mass defect analysis (**Figure S5**) reveals the presence of longer aliphatic carbon chains ($C_xH_yO_{1-5}$) with masses extending to 400-500 Da, indicating polymeric species beyond our computational predictions. These experimental signatures suggest that smaller species identified through the computational framework (such as vinyl-bearing formates and cyclic carbonate precursors) could undergo polymerization reactions during SEI formation to produce larger polymeric networks and highly condensed species. Despite these limitations, our approach provides substantial value by narrowing the structural search space from infinite possibilities to well-defined computational candidates, transforming SEI characterization from empirical guessing into hypothesis-driven investigation guided by electrochemically plausible reaction pathways. Finally, we acknowledge that our simulations do not explicitly incorporate explicit electrode surface effects on reaction pathways. Bin Jassar *et al.* showed that inorganic SEI surfaces can significantly alter reaction barriers and selectivity compared to gas-phase conditions.[76] Despite this limitation and given our uncertainty about the exact structure of the electrode surface, the agreement between our computational findings and experimental verifications highlights the strength of our computational framework.

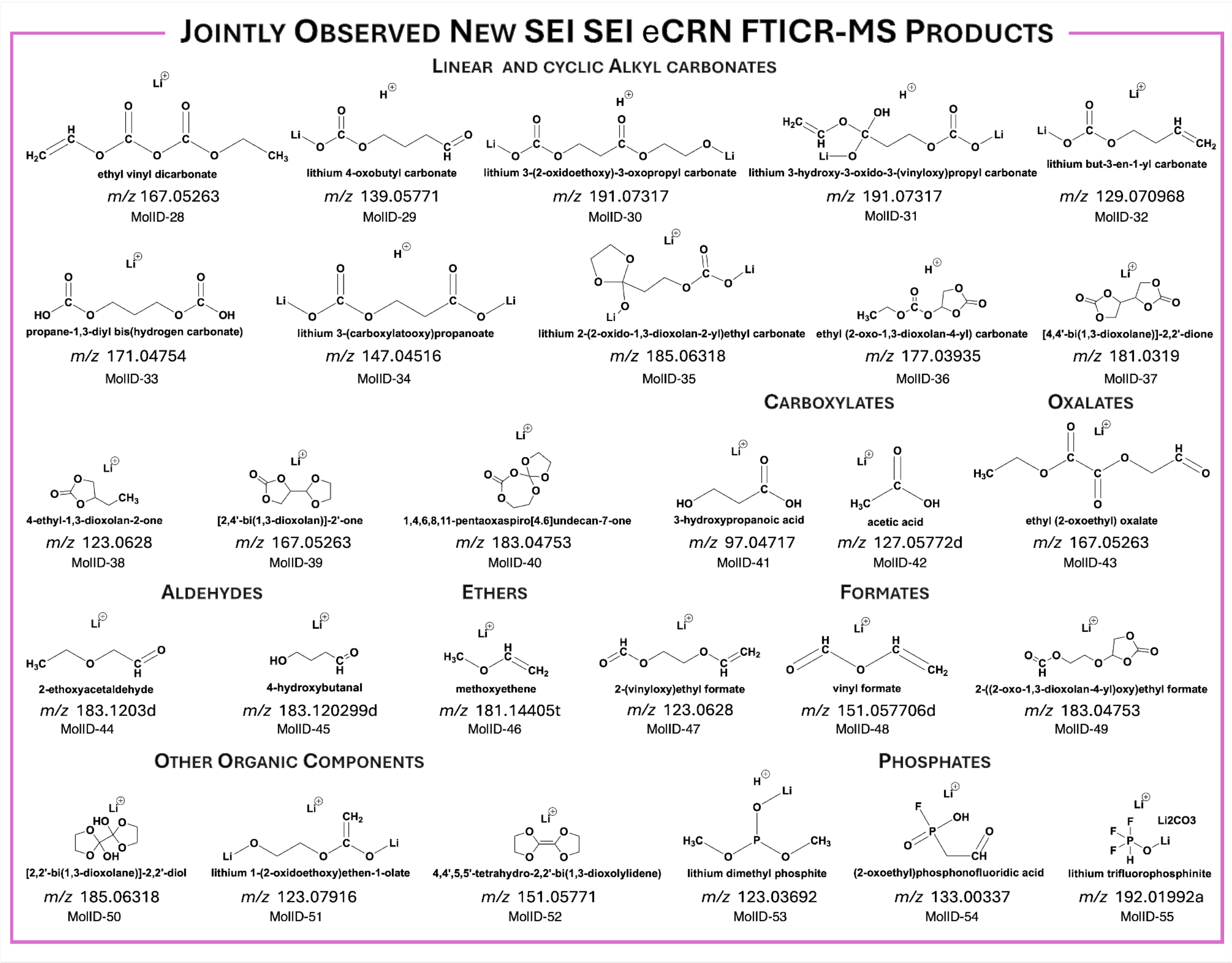

**Figure 4**. Novel SEI species discovered through integrated eCRN, kMC simulations, and LDI-FTICR-MS validation. These molecules, organized by structural motifs, represent a major expansion of SEI chemistry and demonstrate the predictive power of network-based theoretical approaches when coupled with high-resolution mass spectrometry validation. The exact experimental *m/z* of each ion is shown with a hierarchical detection priority: monomer masses are displayed first when detected; if the monomer is not detected but a dimer is found, the dimer mass is shown (marked with "d"); and if neither monomer nor dimer are detected, the trimer mass is displayed (marked with "t"). For MolID-55 with exact mass followed by "a" indicate that this species was only detected with an $Li_2CO_3$ adduct.

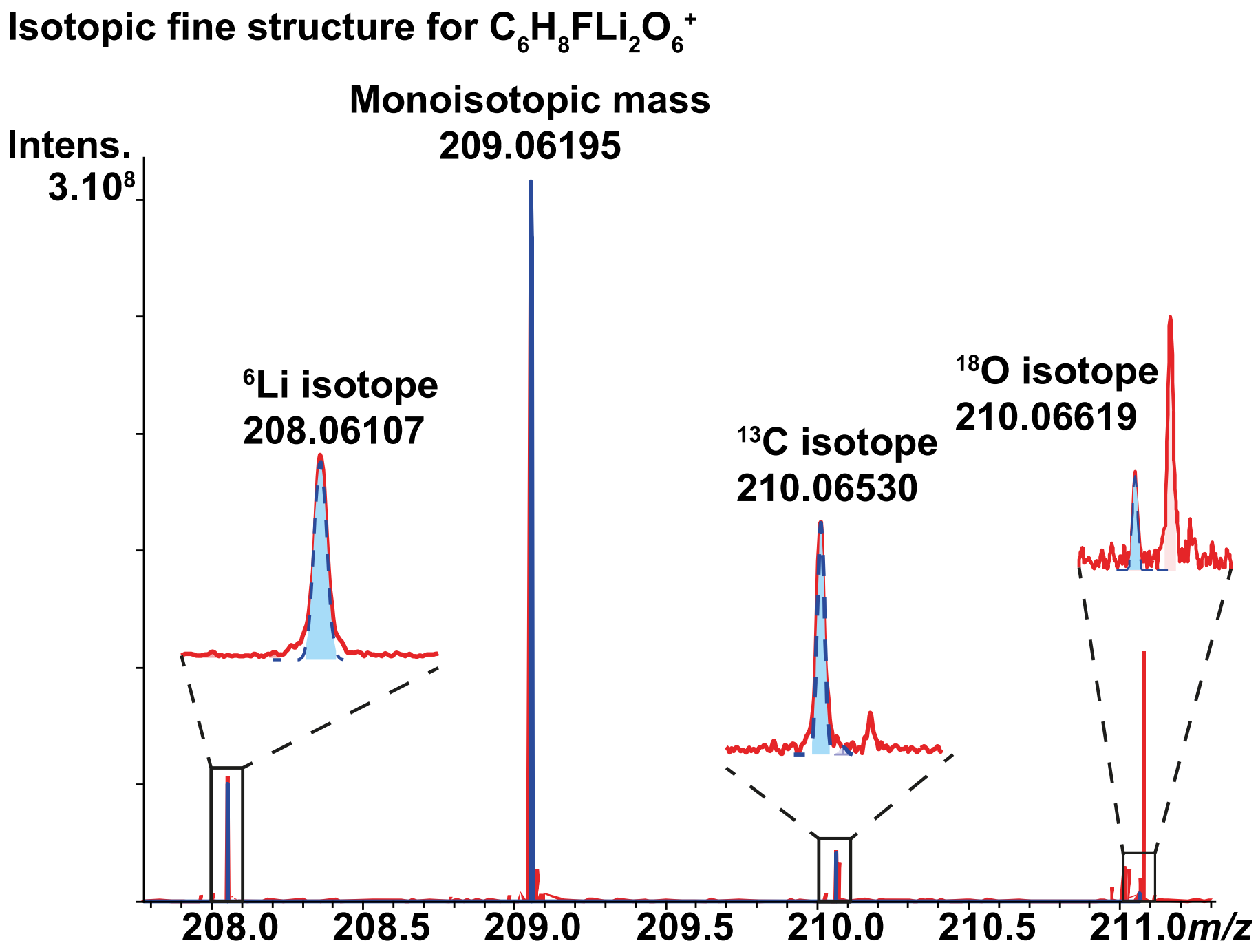

**Figure 5.** MALDI(+) FTICR MS of a negative electrode showing the isotopic fine structure validation of the $C_6H_8FLIi_2O_6^+$ ion (MolID-40 with LiF adduct) with the $^6$Li, $^{13}$C and $^{18}$O isotopes showing perfect matching in mass and intensity between the experimental mass spectrum (in red) and the theoretical profile in blue.

As previously discussed, our framework operates exclusively on thermodynamic data during eCRN generation. Accordingly, the predicted mechanisms should be interpreted as candidate pathways that offer possible mechanistic explanations for the products. Each identified stable species may be formed through the predicted pathways, through alternative pathways not included in the species selection (e.g. by limiting the depth of fragmentation or recombination), or may not form in appreciable amounts due to kinetic hinderance in all possible pathways. The confirmation with LDI-FTICR-MS provided confidence that the proposed species are created in appreciable amounts, eliminating concerns about kinetic hindrance for jointly observed species. However, without kinetic pathway validation, we cannot definitively establish which of the competing mechanistic routes will dominate under specific battery operating conditions, as even thermodynamically favorable processes can be rendered negligible if they encounter prohibitive activation barriers. While resolving the entire eCRN is currently beyond computational capabilities, we kinetically resolved the formation pathways for a representative set of molecules that were jointly observed by experiments and computation (**Figure 6**). All transition states below a 1 eV cutoff were deemed kinetically accessible, a practical threshold that reflects reactions under typical SEI formation conditions (room temperature to moderately elevated potentials). This cutoff also accounts for the evolving nature of the SEI, where species formed via higher barrier pathways are more prevalent in aged, high-cycle interfaces.

The reactants in the included pathways are electrolyte starting components (EMC), previously kinetically resolved species (LiEC, $POF_3$, $Li^{+}EC\text{-}H^{-}$)[29,30,34]. As shown in **Figure 6**, our calculations revealed that $Li^{+}EC\text{-}H^{-}$ (formed by replacing a hydrogen atom on the methylene group of EC with a Li atom) and EMC can undergo at least two distinct reaction pathways, each leading to different product. The first pathway proceeds via nucleophilic attack at the methyl carbon of EMC with an activation barrier of $\Delta G^{\ddagger} = 0.63$ eV, yielding propylene carbonate (MolID-10) and LEC (MolID-9) in a highly thermodynamically favorable process ($\Delta G = -3.46$ eV). The second pathway involves ethyl group transfer, proceeding through a 0.73 eV barrier to form 1,2-butylene carbonate (MolID-46) and LMC (MolID-8), with a reaction energy of -3.25 eV. Furthermore, reactions 3 and 4 represent competing pathways for the consumption of reduced LiEC. Reaction 3, between a ring opened and closed reduced LiEC, exhibits low activation energy barrier of 0.10 eV, with substantial thermodynamic favorability ($\Delta G = -1.44$ eV), suggesting the rapid formation of lithium 2-(2-oxido-1,3-dioxolan-2-yl)ethyl carbonate (MolID-35) under typical battery operating conditions. The competing pathway (reaction 4) between two reduced LiEC molecules for the formation of lithium [2,2'-bi(1,3-dioxolane)]-2,2'-bis(olate) (MolID-51) requires slightly higher kinetic barrier ($\Delta G^{\ddagger} = 0.22$ eV) with more favorable thermodynamic driving force ($\Delta G = -1.79$ eV).

Turning to the $POF_3$-mediated pathways in **Figure 6**, reaction 5a proceeds through a completely barrierless mechanism between $POF_3$ and LiOH ($\Delta G_a^{\ddagger} = 0.00$ eV) to form a $PF_3OHOLi$ intermediate with a reaction energy of -1.52 eV. This intermediate then undergoes a simultaneous bond-breaking bond-formation step with minimal energy barrier of 0.19 eV. The second reaction is thermoneutral ($\Delta G_b = -0.07$ eV), but the reaction energy is calculated for LiF(solv) which is known to deposit LiF(s), a reaction reported to have a high driving force ($\Delta G = -1.17$ eV).[77] The absence of any significant kinetic barriers in this pathway indicates that the formation of LiF and phosphorodifluoridic acid occurs essentially instantaneously upon $POF_3$-LiOH contact. Notably, the formation of LiOH is usually associated with proton-transfer reactions as a result of hydrolysis or solvent decomposition products (see e.g. **Figure S4).**[78] Reaction 6 is a multi-step mechanism leading to the formation of hydrogen fluoride (HF, MolID-26). After the formation of $PF_3OHOLi$ adduct, the O-H bond rotates with an activation barrier of $\Delta G_b^{\ddagger}$ = 0.26 eV and a modest reaction energy of ΔG = -0.29 eV. The subsequent bond-breaking bond-formation step (reaction 6c) to form HF and lithium phosphorodifluoridate (MolID-2) is barrierless ($\Delta G_c^{\ddagger}$ = -0.01 eV), with a reaction energy of -0.66 eV. Overall, the parallel pathways highlighted in **Figure 6** demonstrate a fundamental challenge in SEI prediction, where reactants can access multiple product channels depending on the local electrochemical environment, temperature, and other operating conditions. Consequently, the kinetic competition between these pathways determines the actual product distribution. Moreover, it is highly likely that additional pathways exist for other reactant combinations, particularly those involving radical intermediates or multi-molecular cluster reactions that were not captured in our two-body analysis. For example, the chemical potential of of radical species such as H* and F* are highly sensitive to the local environment and therefore render the reaction energetics context-dependent. This underscores the importance of observing the same products in the mass spectrometry data. In general, we can only claim that these are some of the possible ways that these species are formed. Further work should be done to gain more confidence in which pathways dominate the formation of specific species. This would require comprehensive kinetic modeling that accounts for pathway competition, intermediate stability, and the influence of local electrochemical potential on reaction barriers. Only through this more rigorous analysis may we gain the knowledge necessary to make mechanistically-informed decisions regarding changes to the electrolyte's initial composition.

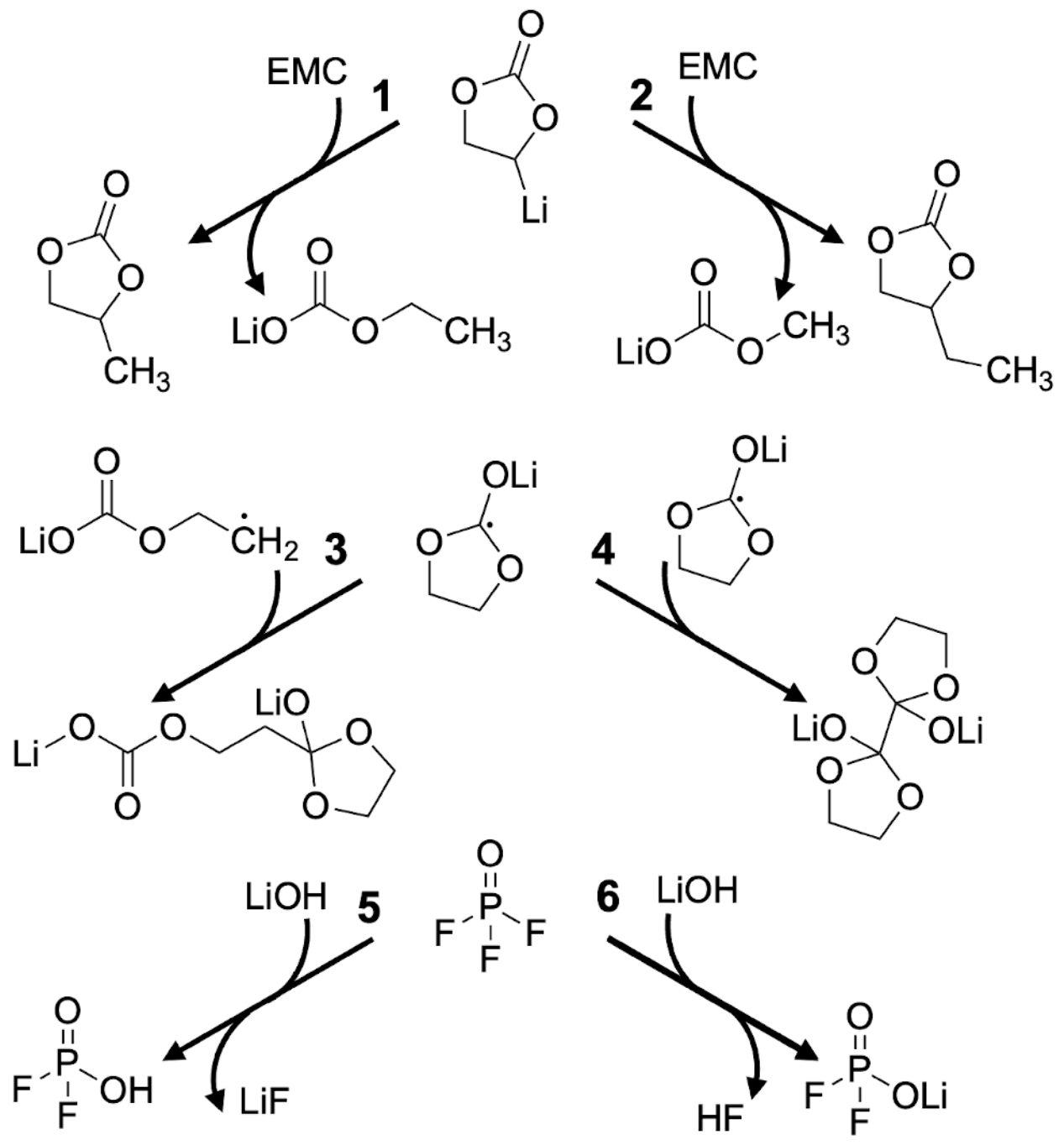

| Reaction | $\Delta G^{\ddagger}$ [eV] | $\Delta G$ [eV] |
|---|---|---|
| 1 | 0.63 | −3.46 |
| 2 | 0.73 | −3.25 |
| 3 | 0.10 | −1.44 |
| 4 | 0.22 | −1.79 |
| 5a | 0.00 | −1.52 |
| 5b | 0.19 | −0.07 |
| 6a | 0.00 | −1.27 |
| 6b | 0.26 | −0.29 |
| 6c | −0.01 | −0.66 |

**Figure 6**. Possible routes for the formation of a representative set of molecules that were jointly observed by experiments and computation. For multi-step reactions such as reaction 5 and 6, a, b, and c represent sequential elementary steps (step 1, step 2, and step 3, respectively) with their corresponding activation energy barriers ($\Delta G^{\ddagger}$) and reaction free energies ($\Delta G$) provided. Graphical snapshots of full energy diagram and relaxed structures (e.g. TSs) are depicted in **Figures S7-S9**.

Looking forward, the integration of machine learning interatomic potentials (MLIPs) with species selection represents an opportunity to overcome the current computational bottlenecks that limit kinetic analysis to small subsets of reactions. While our detailed kinetic refinement has revealed critical mechanistic insights for key pathways, the generated reaction network contains millions of potential reactions that remain kinetically unresolved. The development of MLIPs trained on high-quality DFT transition state data could enable rapid estimation of activation barriers across the entire reaction network, finally bridging the gap between thermodynamic feasibility and kinetic accessibility that currently limits eCRN predictive power. Recent advances in universal MLIPs and automated transition state search algorithms suggest that such an approach could computationally screen millions of elementary steps at near-DFT accuracy within practical timeframes.[79,80] Alternatively, physics-based semi-empirical methods offers orders of magnitude computational speedup over DFT.[81] However, it is essential to note that these methods require ongoing validation, reparameterization, and careful benchmarking due to their reliance on approximations and empirical data rather than purely first-principles calculations. By leveraging automated eCRN generation and stochastic network analysis, future frameworks could uncover new reaction pathways, identify key intermediates, and improve the reliability of kinetic models used in a variety of applications from reactor design, emissions control, electrochemical devices to plastics degradation and fuel optimization.

**Conclusion:**

In this work, we provided a comprehensive molecular-level characterization of the SEI formed on graphite in a carbonate-based Li-ion battery electrolyte through the integration of ultra-high-resolution LDI-FTICR-MS, data-driven electrochemical reaction networks, and stochastic simulations. The synergetic combination of these simulations and high-resolution experimental techniques leads to the identification of new molecular species with high confidence. We generated the largest and most comprehensive eCRN for SEI formation in LIBs, consisting of over 10,000 species and 209 million reactions. Stochastic network analysis yielded many known SEI components as well as novel species that had not previously been proposed. These newly identified molecules were validated with LDI-FTICR-MS and spanned diverse structural motifs including fluorinated organophosphate-carbonates, cyclic and bicyclic alkyl carbonates, carboxylates, formats, diols, formats, oxalates, dioxolanes, and ethers that collectively expand our understanding of SEI chemical diversity. Kinetic analysis of formation pathways for a set of representative molecules revealed kinetically feasible routes with activation energy barriers, ranging from completely barrierless to less than 1.0 eV. Together, this work lays a foundation for the rational design of electrolytes, offering a strategic pathway toward more stable, efficient, and longer-lasting lithium-based energy storage systems.

**Supporting Information.** Formation cycle of NCM/Graphite pouch cells; First-cycle capacity data; FTICR MS of the four negatives graphite electrodes; Species and reactions filters; FTICR MS molecular assignment for all novel species; Structural comparison of isomers with similar *m/z.*

**Acknowledgement:**

This work was supported by a research grant from TotalEnergies to the University of California, Berkeley. M.A. acknowledge funding support from the Bakar Institute of Digital Materials for the Planet Postdoctoral Fellowship. The research was performed using computational resources sponsored by the Department of Energy's Office of Critical Minerals and Energy innovation and located at the National Laboratory of the Rockies. This research also used the Lawrencium computational cluster resource provided by the IT Division at the Lawrence Berkeley National Laboratory (supported by the Director, Office of Science, Office of Basic Energy Sciences, of the U.S. Department of Energy under Contract No. DE-AC02-05CH11231). Schrödinger, Inc. provided access to Jaguar and AutoTS software. C.A. acknowledge financial support from Labex SynOrg (ANR-11-LABX-0029) and the CNRS research infrastructure Infranalytics (FR2054) for instrument time.

# Supporting Information

# for

# Identification of Solid-Electrolyte Interphase Species by Joint Characterization of Li-ion Battery Chemistry by Mass Spectrometry and Electro-Chemical Reaction Networks

Mona Abdelgaid,[1,2] Oliver Hvidsten,[2,3] Théo Sombret,[4-7] Egon Kherchiche,[4-7] Julien Maillard,[5,7] Antonin Gajan,[6] Patrick Bernard,[6] Kamila Kazmierczak,[8] Mauricio Araya-Polo,[9] Germain Salvato Vallverdu,[7,10] Carlos Afonso,[4,7] Pierre Giusti,[4,5,7] Kristin A. Persson*[1-3]

11. Bakar Institute of Digital Materials for the Planet, University of California at Berkeley, Berkeley, CA, 94720, United States.
12. Materials Science Division, Lawrence Berkeley National Laboratory, Berkeley, CA, 94720, United States.
13. Department of Materials Science and Engineering, University of California at Berkeley, Berkeley, CA, 94720, United States.
14. Université de Rouen Normandie, INSA Rouen Normandie, Université de Caen Normandie, ENSICAEN, CNRS, Institut CARMeN UMR 6064, F-76821 Mont-Saint-Aignan Cedex, France.
15. TotalEnergies OneTech, TotalEnergies Research & Technology Gonfreville, BP 27, 76700 Harfleur, France.
16. Saft, Corporate Research, 33074 Bordeaux, France.
17. International Joint Laboratory-Complex Matrices Molecular Characterization (iC2MC), TRTG, BP 27, 76700 Harfleur, France.
18. TotalEnergies OneTech Belgium, 7181 Seneffe, Belgium.
19. TotalEnergies EP Research & Technology, Houston, TX, 77002, United States.
20. Universite de Pau et des Pays de l'Adour, CNRS, IPREM, 64000 Pau, France.

*Corresponding author

E-mail: kapersson@lbl.gov

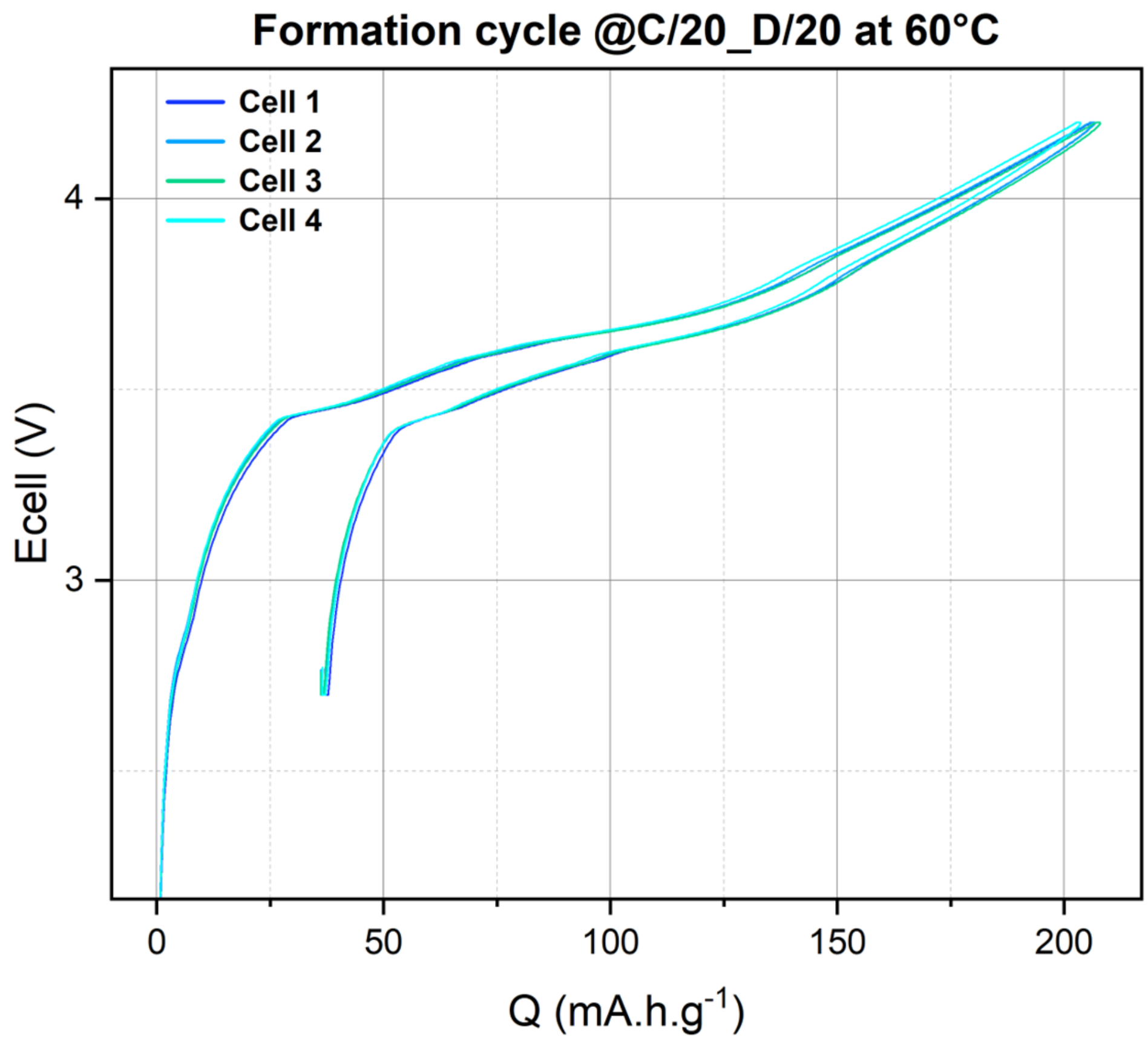

**Figure S1.** Formation cycle of NCM/Graphite pouch cells between 2.70 and 4.20 V using currents corresponding to C/20 at 60 °C with a reference electrolyte: 1 M $LiPF_6$ in EC/EMC (30/70 v).

**Table S1.** First-cycle capacity data: initial charge ($Q_{1st\ charge}$), irreversible loss ($Q_{irreversible}$), and reversible capacity ($Q_{reversible}$) for the four cells.

| $Q_{1st\ charge}$ ($mAh.g^{-1}$) | $Q_{irreversible}$ ($mAh.g^{-1}$) | $Q_{reversible}$ ($mAh.g^{-1}$) |
|---|---|---|
| 207 | 37 | 170 |
| 206 | 36 | 170 |
| 208 | 36 | 172 |
| 204 | 36 | 168 |

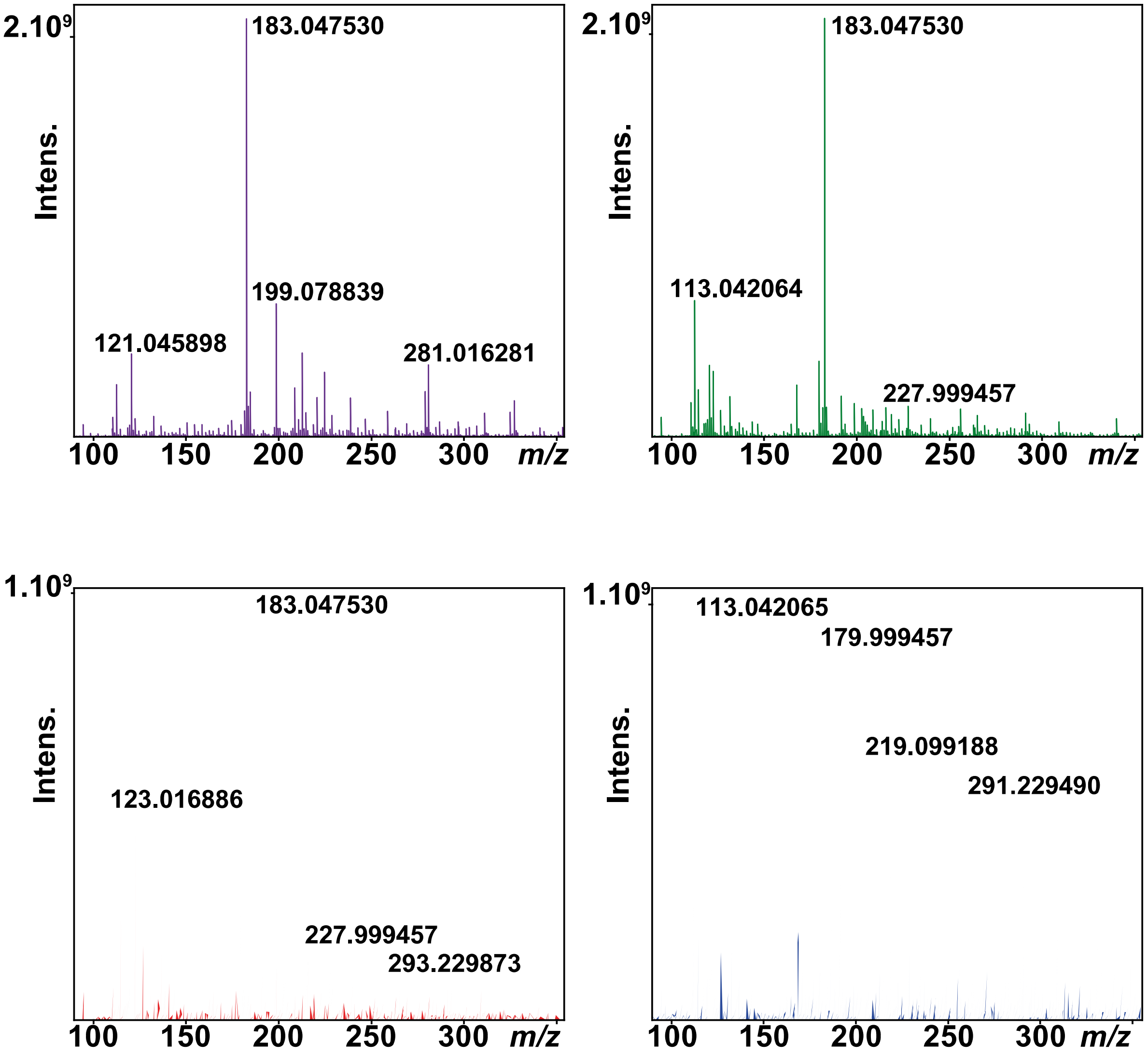

**Figure S2.** MALDI(+) FTICR MS of the four negatives graphite electrodes on the *m/z* 90 – *m/z* 350 range.

## Species Filters:

1. Metal Non-Cation Filter:
   - Remove species with Li atom having an NBO partial charge < 0.1.
2. Molecule Not Connected:
   - Remove species lacking a connected full molecule graph (both covalent and coordinate bonding).
3. Metal-Centric Complex:
   - Remove species where connectivity is solely due to a metal ion (e.g., A — $Li^+$ — B).

## Reaction Filters

1. Endergonic reactions exhibiting $\Delta G > 0$.
2. Reactions featuring charge change $|\Delta q| > 1$.
3. Redox reactions containing multiple reactants or multiple products.
4. Unimolecular dissociative redox reactions where $|\Delta q| > 0$ and covalent bond formation or breakage occurs.
5. Reactions involving more than two reactants or products.
6. Reactions containing spectator species (components that remain unchanged throughout the reaction), such as A + B → A + C.
7. Reactions featuring more than two bond modifications.
8. Reactions where dual bonds form concurrently or dual bonds break concurrently.
9. Reactions where covalent bond alterations coincide with metal ion coordination or decoordination (excluding reactions where metal ions maintain coordination while modifying their coordinate bonds, which remain permissible).

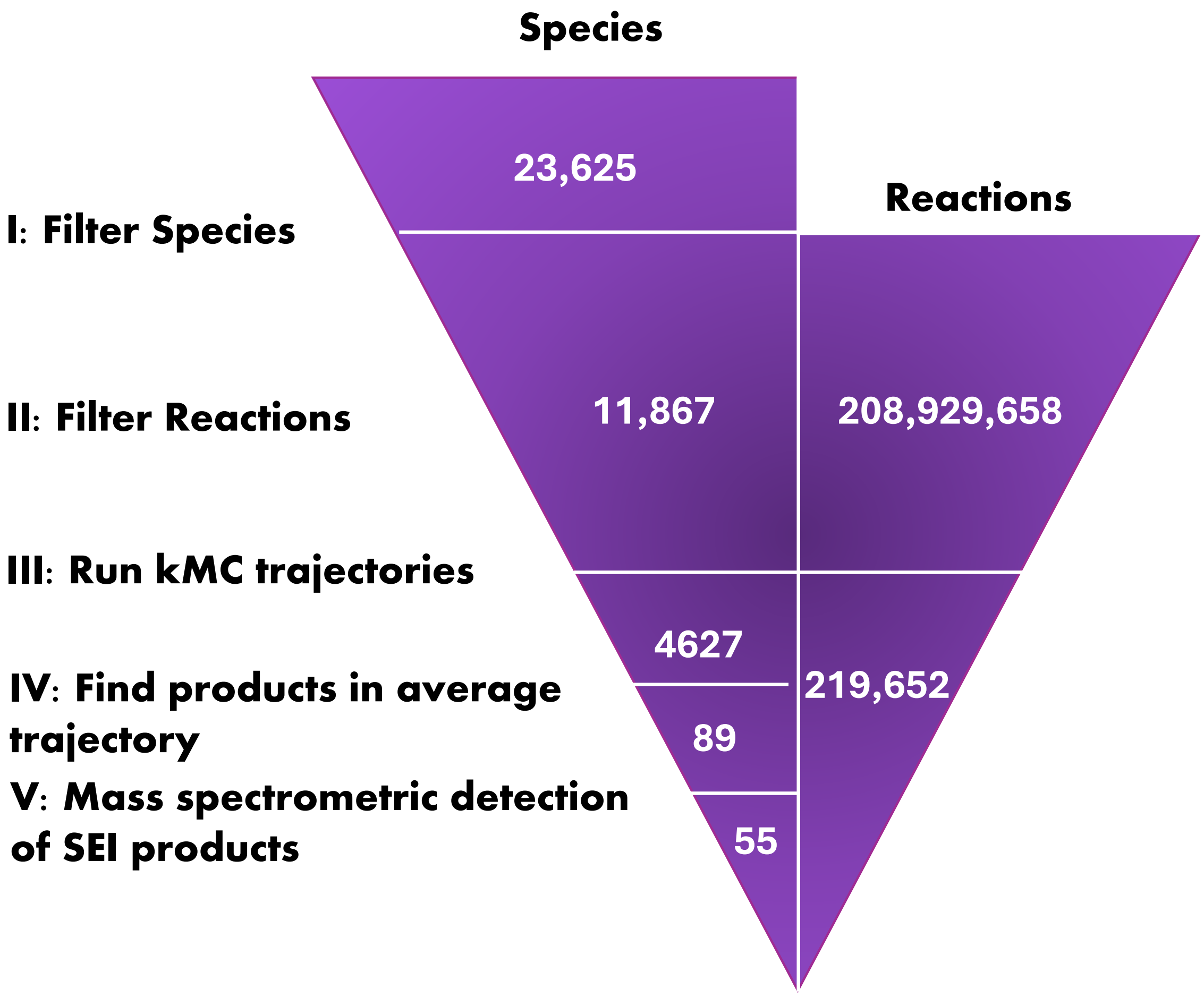

**Figure S3.** Electrochemical reaction network construction and stochastic analysis for SEI formation yields a systematic reduction of complexity.

**Table S2.** MALDI(+) FTICR MS molecular assignment with theoretical exact mass, experimental measurement, and relative error for all the new molecular formula found on the mass spectra. The identification parameters, after calibration on the carbon clusters, were setup as followed: allowed atoms $C_n$, $H_{40}$, $O_n$, $N_0$, $F_{15}$, $Li_5$, $P_5$ with a tolerance of 0.5 ppm.

| Ion molecular formula | Theoretical *m/z* | Experimental *m/z* | Relative error (ppm) |
|---|---|---|---|
| $C_2H_6Li_2O_3P^+$ | 123.036914 | 123.036916 | 0.01 |
| $C_6H_9O_6^+$ | 177.039364 | 177.039354 | −0.06 |
| $C_6H_{10}LiO_6^+$ | 185.063193 | 185.063180 | −0.07 |
| $C_6H_8Li_3O_6^+$ | 197.079549 | 197.079580 | 0.15 |
| $C_5H_8LiO_3^+$ | 123.062799 | 123.062802 | 0.02 |
| $C_6H_6LiO_6^+$ | 181.031893 | 181.031895 | 0.01 |
| $C_6H_8LiO_5^+$ | 167.052628 | 167.052625 | −0.02 |
| $C_6H_8LiO_6^+$ | 183.047543 | 183.047530 | −0.07 |
| $C_5H_8LiO_4^+$ | 139.057714 | 139.057712 | −0.01 |
| $C_5H_7Li_2O_6^+$ | 177.055721 | 177.055732 | 0.06 |
| $C_5H_8LiO_6^+$ | 171.047543 | 171.047535 | −0.05 |
| $C_4H_5Li_2O_5^+$ | 147.045157 | 147.045153 | −0.02 |
| $C_3H_6LiO_3^+$ | 97.047149 | 97.047171 | 0.23 |
| $C_5H_9O_3^+$ | 117.054621 | 117.054623 | 0.02 |
| $C_6H_8LiO_4^+$ | 151.057714 | 151.057711 | −0.02 |
| $C_4H_6Li_3O_3^+$ | 123.079156 | 123.079164 | 0.07 |
| $C_{10}H_{16}LiO_8^+$ | 271.099972 | 271.099975 | 0.01 |
| $C_6H_{12}LiO_6^+$ | 187.078843 | 187.078845 | 0.01 |
| $C_6H_8Li_3O_6^+$ | 197.079550 | 197.079580 | 0.15 |
| $C_4H_8LiO_4^+$ | 127.057714 | 127.057717 | 0.03 |
| $C_8H_{16}LiO_4^+$ | 183.120314 | 183.120309 | −0.03 |
| $C_8H_{16}LiO_6^+$ | 215.110143 | 215.110142 | −0.01 |
| $C_{12}H_{24}LiO_6^+$ | 271.172743 | 271.172743 | 0.00 |
| $C_9H_{18}LiO_3^+$ | 181.141049 | 181.141046 | −0.02 |
| $C_5FH_8Li_2O_5^+$ | 181.067035 | 181.067041 | 0.03 |
| $C_6FH_8Li_2O_6^+$ | 209.061949 | 209.061947 | −0.04 |
| $C_6H_9Li_2O_6^+$ | 191.071371 | 191.071371 | 0.00 |
| $C_6H_8Li_3O_6^+$ | 197.079550 | 197.079580 | 0.15 |
| $C_5FH_8Li_2O_5^+$ | 181.067035 | 181.067041 | 0.03 |
| $C_6H_9Li_2O_7^+$ | 207.066286 | 207.066265 | −0.10 |
| $C_6FH_{10}Li_2O_6^+$ | 211.077599 | 211.077607 | 0.04 |
| $C_6H_7Li_4O_9^+$ | 251.072472 | 251.072461 | −0.04 |
| $C_6FH_{10}Li_2O_6^+$ | 211.077599 | 211.077607 | 0.04 |
| $C_5FH_6Li_2O_5^+$ | 179.051385 | 179.051377 | −0.04 |
| $C_5H_9Li_2O_5^+$ | 163.076457 | 163.076442 | −0.09 |
| $C_5H_8Li_3O_5^+$ | 169.084635 | 169.084615 | −0.12 |
| $C_4FH_8Li_2O_2^+$ | 121.082291 | 121.082244 | −0.39 |
| $C_4H_7Li_2O_4^+$ | 133.065892 | 133.065887 | −0.04 |
| $C_4H_6Li_3O_4^+$ | 139.074070 | 139.074073 | 0.02 |
| $C_7H_9Li_2O_7^+$ | 219.066286 | 219.066278 | −0.04 |
| $C_6FH_8Li_2O_4^+$ | 177.072120 | 177.072101 | −0.11 |
| $C_6H_9Li_2O_6^+$ | 191.071371 | 191.071372 | 0.00 |
| $C_6FH_{10}Li_2O_6^+$ | 211.077599 | 211.077607 | 0.04 |
| $C_5H_8Li_3O_6^+$ | 185.079549 | 185.079546 | −0.02 |
| $C_5H_6Li_5O_6^+$ | 197.095907 | 197.095903 | −0.02 |
| $C_4FH_8Li_2O_3^+$ | 137.077206 | 137.077212 | 0.05 |
| $C_6FH_8Li_2O_6^+$ | 209.061949 | 209.061947 | −0.01 |

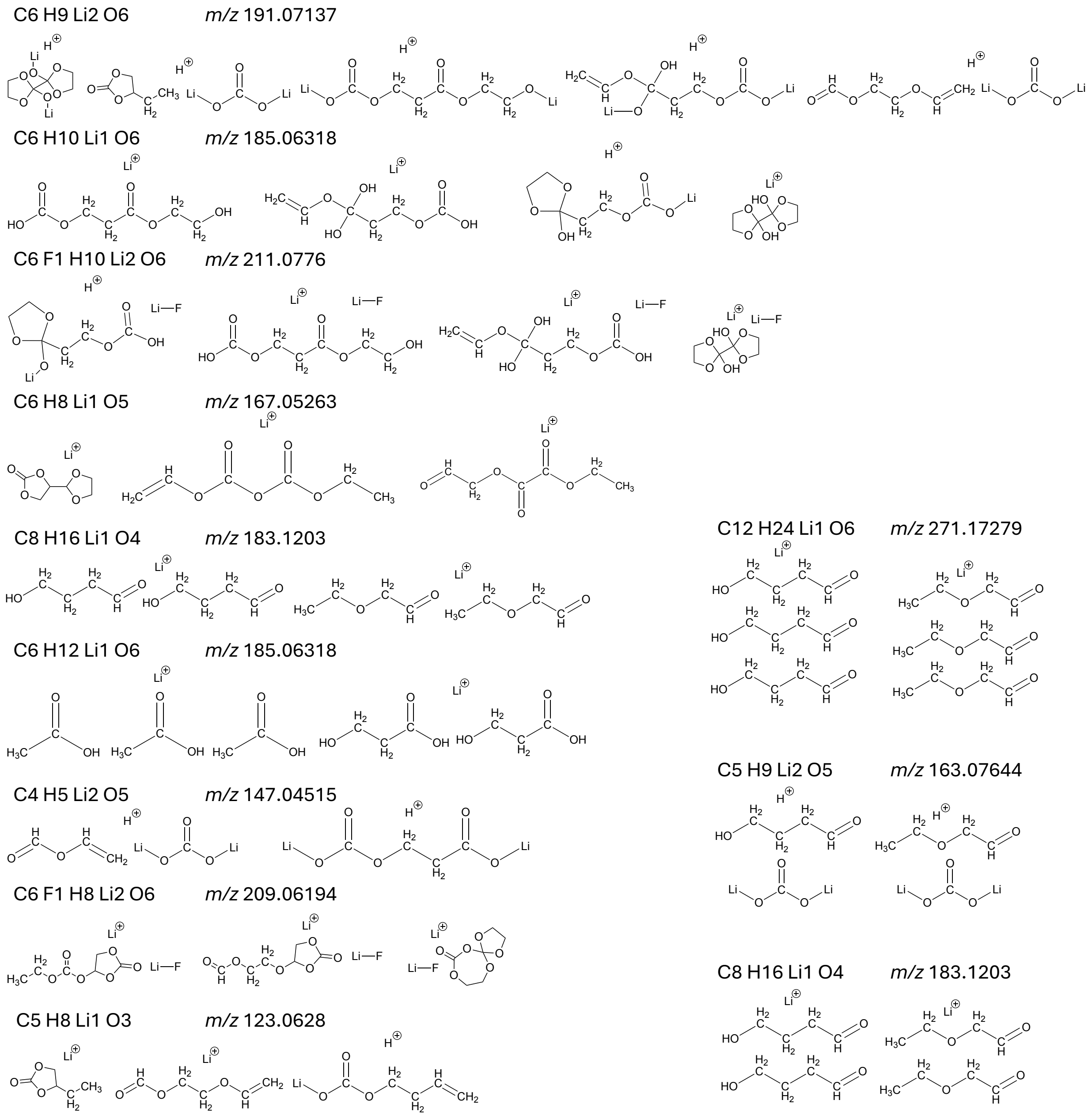

**Figure S4.** Structural comparison of isomers with similar *m/z*.

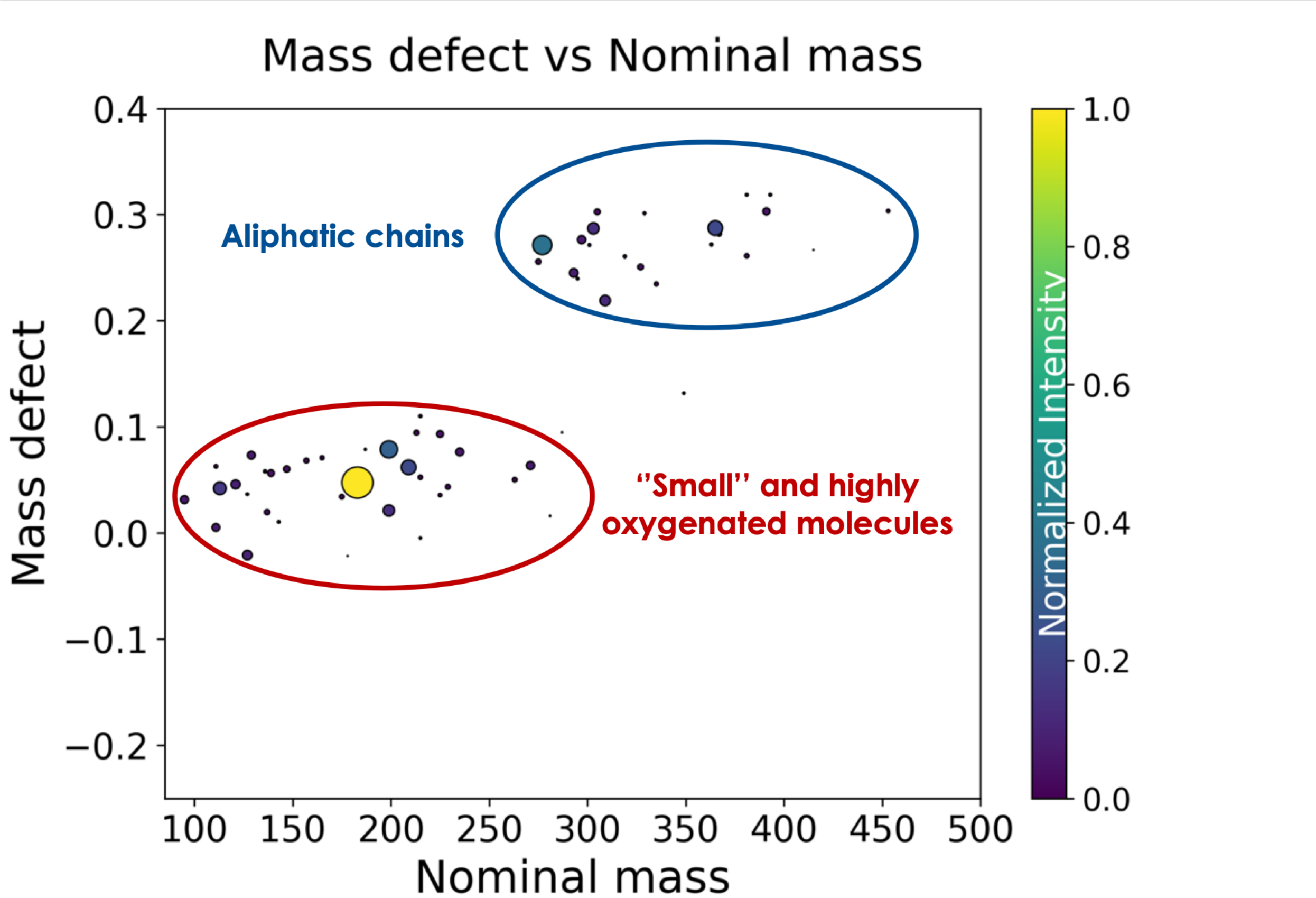

**Figure S5**. 2D mass spectrum representation: nominal mass vs. mass defect mapping of SEI species after the formation cycle. The mass defect is calculated as the difference between exact mass and nominal mass for each detected species. The size and color intensity of each data point correspond to the relative ion intensity in the mass spectrum.

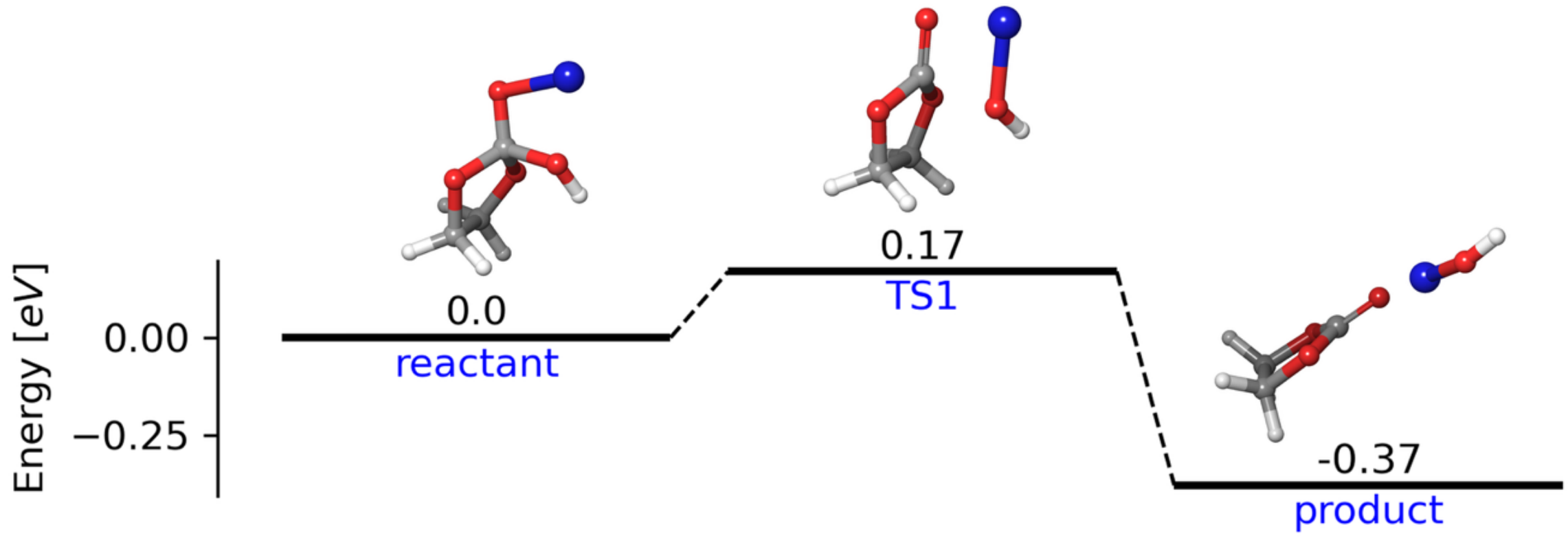

**Figure S6**. Energy diagram for the formation of LiOH.

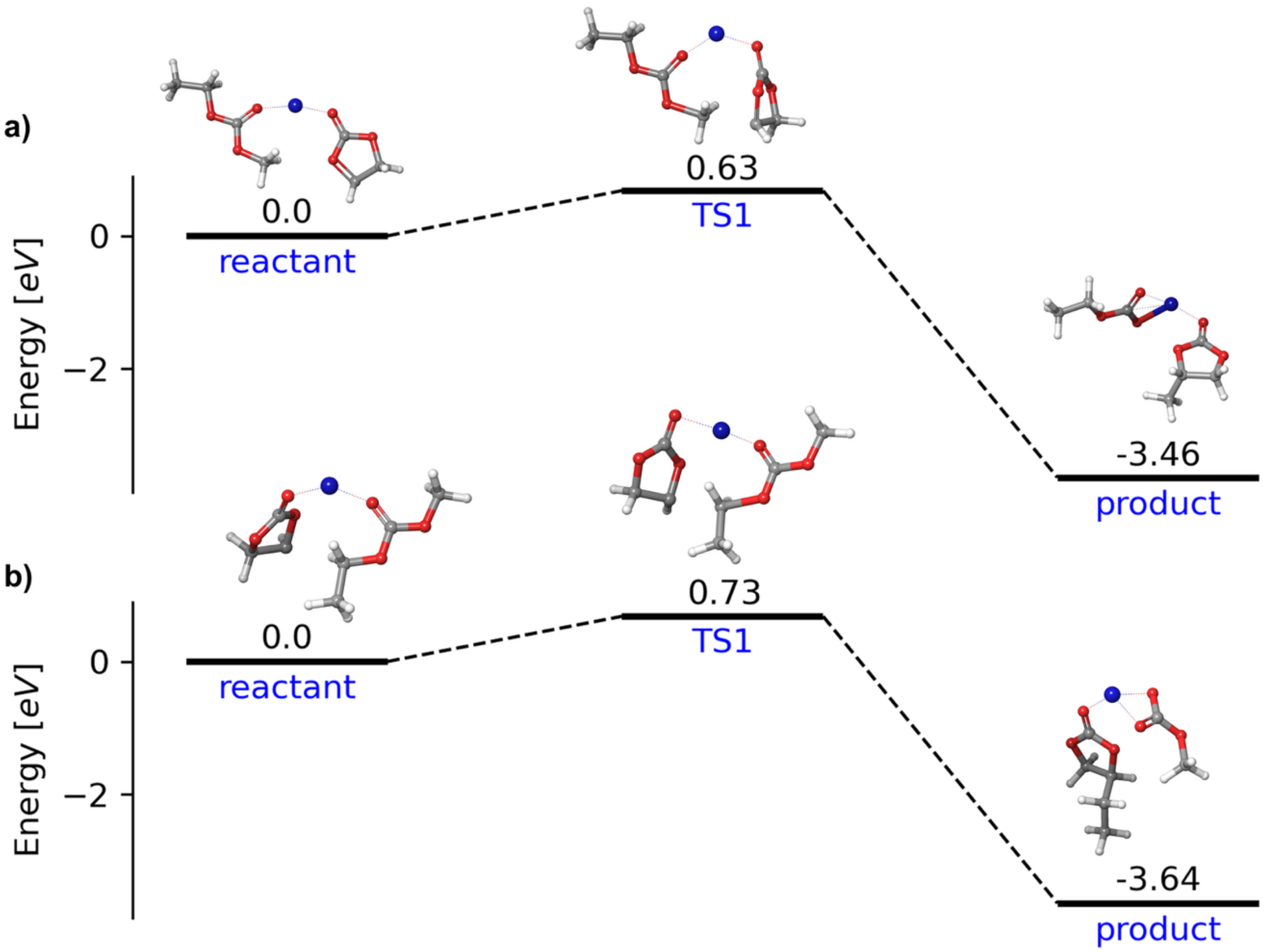

**Figure S7**. Energy diagram for the formation of (a) propylene carbonate (MolID-10) and lithium ethyl carbonate (MolID-9) and (b) 1,2-butylene carbonate (MolID-38) and lithium methyl carbonate (MolID-8).

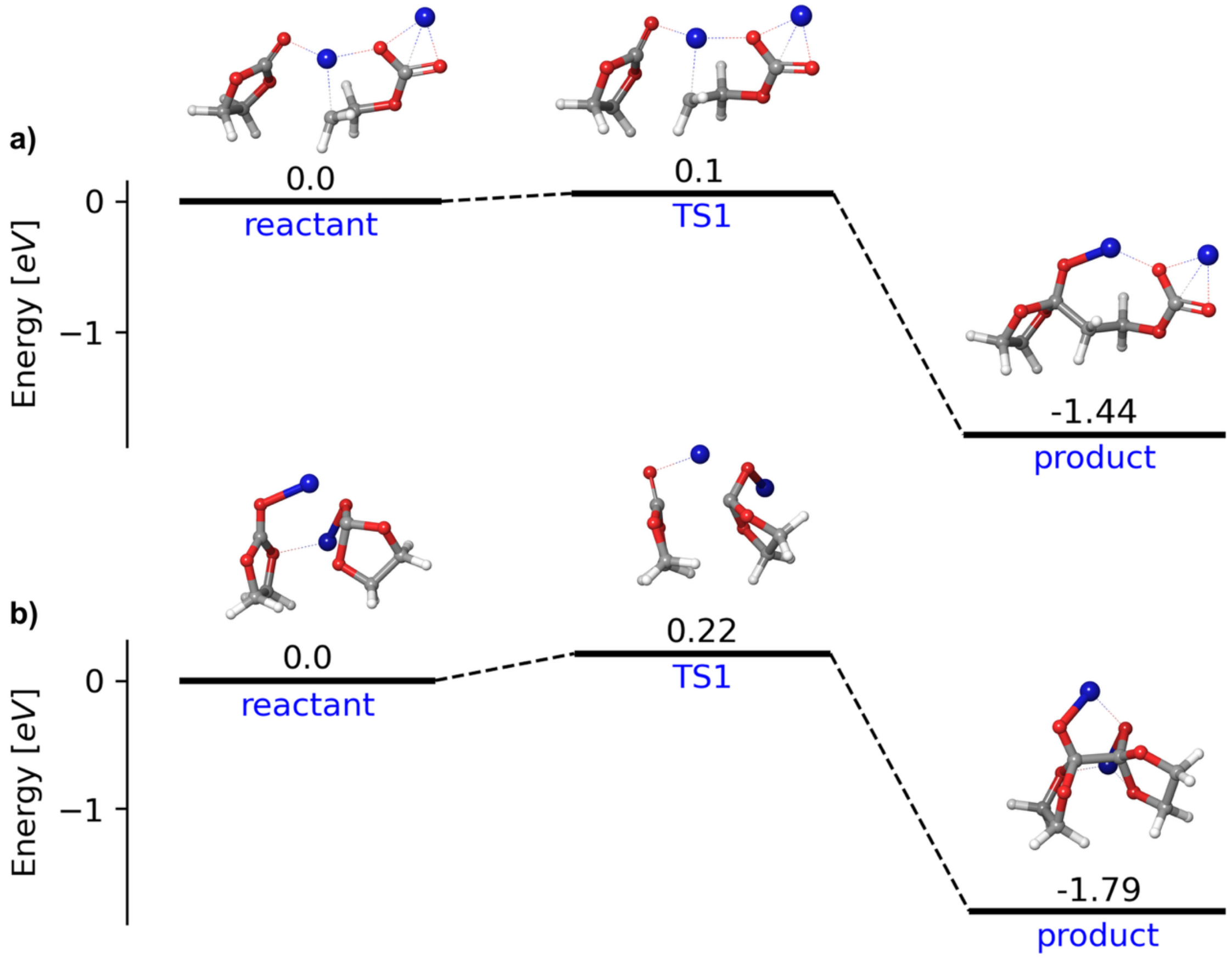

**Figure S8**. Energy diagram for the formation of (a) lithium 2-(2-oxido-1,3-dioxolan-2-yl)ethyl carbonate (MolID-35) and (b) lithium [2,2'-bi(1,3-dioxolane)]-2,2'-bis(olate) (MolID-50).

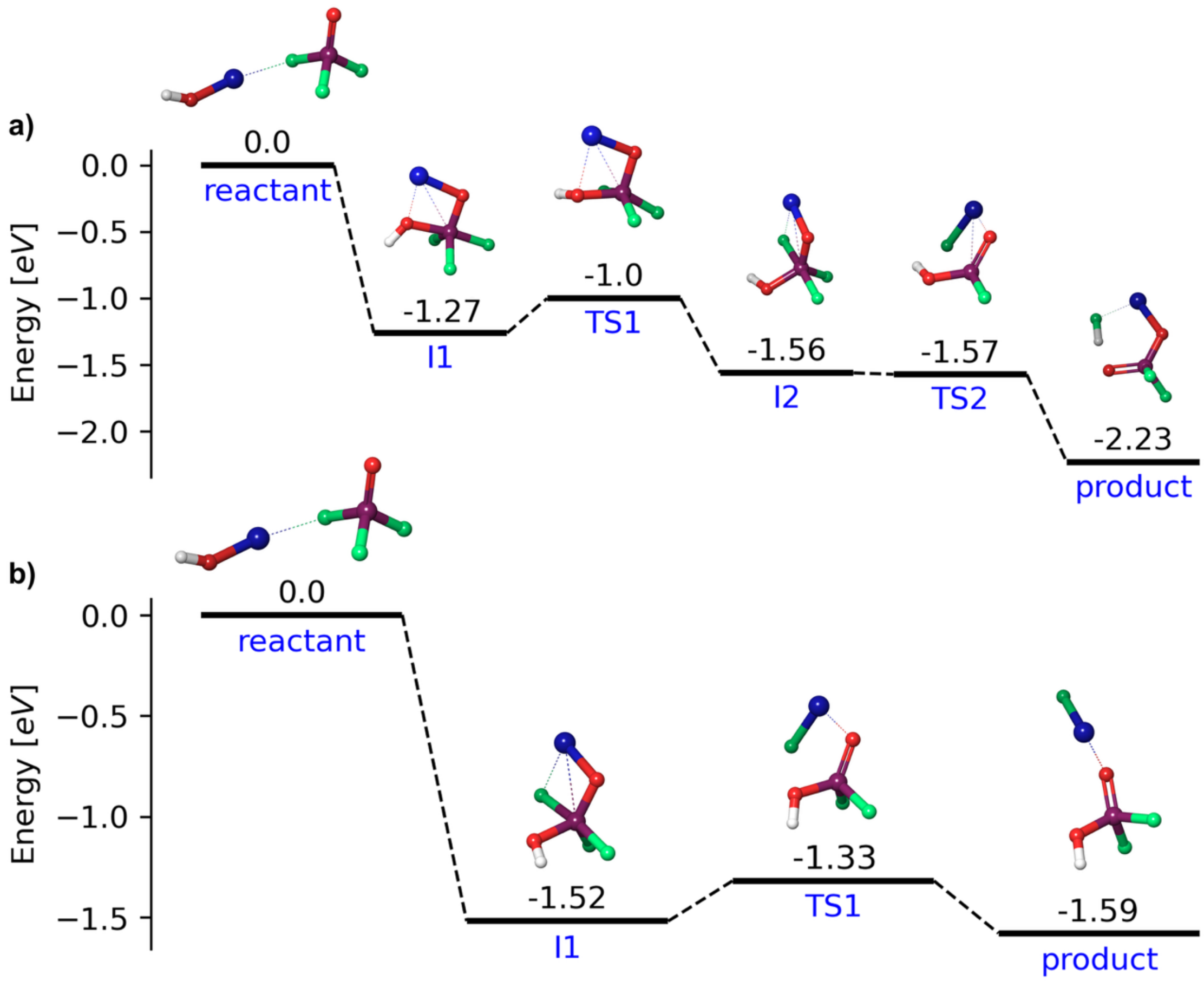

**Figure S9**. Energy diagram for the formation of (a) $POF_2OLi$ (MolID-2) and HF (MolID-26) and (b) $POF_2OH$ and LiF (MolID-23).